\pdfoutput=1
\documentclass{jfm}
\usepackage{}

\usepackage{graphicx,epsfig}
\usepackage{mathrsfs}
\usepackage{amsmath}
\usepackage{amsfonts}
\usepackage{amssymb}
\usepackage{color}

\ifCUPmtlplainloaded \else
  \checkfont{eurm10}
  \iffontfound
    \IfFileExists{upmath.sty}
      {\typeout{^^JFound AMS Euler Roman fonts on the system,
                   using the 'upmath' package.^^J}%
       \usepackage{upmath}}
      {\typeout{^^JFound AMS Euler Roman fonts on the system, but you
                   dont seem to have the}%
       \typeout{'upmath' package installed. JFM.cls can take advantage
                 of these fonts,^^Jif you use 'upmath' package.^^J}%
      }
  \else
  \fi
\fi

\ifCUPmtlplainloaded \else
  \checkfont{msam10}
  \iffontfound
    \IfFileExists{amssymb.sty}
      {\typeout{^^JFound AMS Symbol fonts on the system, using the
                'amssymb' package.^^J}%
       \usepackage{amssymb}%
       \let\le=\leqslant  \let\leq=\leqslant
       \let\ge=\geqslant  \let\geq=\geqslant
      }{}
  \fi
\fi

\ifCUPmtlplainloaded \else
  \IfFileExists{amsbsy.sty}
    {\typeout{^^JFound the 'amsbsy' package on the system, using it.^^J}%
     \usepackage{amsbsy}}
    {\providecommand\boldsymbol[1]{\mbox{\boldmath $##1$}}}
\fi

\newsavebox{\astrutbox}
\sbox{\astrutbox}{\rule[-5pt]{0pt}{20pt}}

\newcommand\etal{\mbox{\textit{et al.}}}

\newcommand\beq{\begin{equation}}
\newcommand\eeq{\end{equation}}
\newcommand\beqa{\begin{eqnarray}}
\newcommand\eeqa{\end{eqnarray}}

\newcommand{\bphi}{\mbox{\boldmath$\phi$}}

\newcommand{\bnab}{\mbox{\boldmath$\nabla$}}
\newcommand{\bnu}{\mbox{\boldmath$\nu$}}
\newcommand{\eps}{\varepsilon}

\def\bnab{\mbox{\boldmath $ \nabla$}}

\def\half {{\textstyle{1 \over 2}}}

\newcommand{\bx}{\mathbf{x}}

\newcommand{\bhz}{\mathbf{\hat{z}}}

\newcommand{\bu}{\mathbf{u}}

\newcommand{\opL}{{\mathscr L}}
\newcommand{\T}{{\cal T}}
\def\N{\mbox{\boldmath${\cal N}$}}
\def\H{{\cal H}}
\def\<{\langle}
\def\>{\rangle}

\title[]{Exhausting the background approach for bounding the heat transport in Rayleigh-B\'enard convection}

\author[Z. Ding and R. R. Kerswell]%
{Zijing Ding$^1$\thanks{z.ding@damtp.cam.ac.uk}\ns
and\ns Rich R. Kerswell$^1$\thanks{r.r.kerswell@damtp.cam.ac.uk}}

\affiliation{Department of Applied Mathematics and Theoretical Physics, Centre for Mathematical Sciences, University of Cambridge, Cambridge, CB3 0WA, UK.
}

\date{?; revised ?; accepted ?.}

\begin{document}

\maketitle

\begin{abstract}

We revisit the optimal heat transport problem for Rayleigh-B\'enard convection in which a rigorous upper bound on the Nusselt number, $Nu$, is sought as a function of the Rayleigh number $Ra$. Concentrating on the 2-dimensional problem with stress-free boundary conditions, we impose the full heat equation as a constraint for the bound using a novel 2-dimensional background approach thereby complementing the `wall-to-wall' approach of Hassanzadeh \etal \,(\emph{J. Fluid Mech.} \textbf{751}, 627-662, 2014). Imposing the same symmetry on the problem, we find correspondence with their result for $Ra \leq Ra_c:=4468.8$ but, beyond that, the optimal fields complexify to produce a higher bound.  This bound approaches that by a 1-dimensional background field as the length of computational domain $L\rightarrow\infty$. On lifting the imposed symmetry, the optimal 2-dimensional temperature background field reverts back  to being 1-dimensional  giving the best bound $Nu\le 0.055Ra^{1/2}$ compared to $Nu \le 0.026Ra^{1/2}$ in the non-slip case.
We then show via an inductive bifurcation analysis that imposing the full time-averaged Boussinesq equations as constraints (by introducing 2-dimensional temperature {\em and} velocity background fields) is also unable to lower this bound. This then exhausts the background approach for  the 2-dimensional (and by extension 3-dimensional) Rayleigh-Benard problem with the bound remaining stubbornly $Ra^{1/2}$ while data seems more to scale like $Ra^{1/3}$ for large $Ra$.
Finally, we show that adding a velocity background field to the formulation of Wen \etal\, (\emph{Phys. Rev. E.} \textbf{92}, 043012, 2015), which is able to use an extra vorticity constraint  due to the stress-free condition to lower the bound to $ Nu \le O(Ra^{5/12})$, also fails to improve the bound.

\end{abstract}

\begin{keywords}
Upper bound, Rayleigh-B\'enard convection
\end{keywords}

%
%
\section{Introduction}

In this paper we consider the fundamental problem of assessing how the heat flux per unit area behaves as a function of the Rayleigh number, $Ra$,  in Rayleigh-Benard convection where a layer of fluid is heated from below and cooled from above. This situation  is ubiquitous in Nature and consequently the focus of a huge body of ongoing research work (e.g. Ahlers et  al.  2009). The particular focus here is the use of variational methods which seek an upper bound on the heat flux in the hope that this bound will capture the correct high-$Ra$ scaling for turbulent convection. This approach involves constructing an optimisation problem constrained by information  gleaned from the governing equations. Inevitably, the constraints actually imposed form a strict subset of those implied by the governing equations so that any maximum which emerges is an upper bound on what can actually be realised.  This approach has its roots in the work of Malkus (1954) who hypothesized that the fluid selects the flow state from all those possible states which maximises the heat  transport. The subsequent mathematical formulation by Howard (1963) and Busse (1969) was as a maximization problem (see the early reviews by Howard (1972) and Busse (1978)\,).  In the 1990s, an alternative complementary approach - the background  method - was introduced by Doering \& Constantin (1992,1994,1995,1996) which takes the form of a {\em minimization} problem. This has the considerable advantage that even a trial solution can yield an upper bound which, experience seems to indicate, yields the same scaling as the proper optimal (e.g. in shear flow and convection see Doering \& Constantin 1992,1996 respectively compared to Plasting \& Kerswell 2003, hereafter PK03).

In both approaches, however, the outstanding challenge has been to add further dynamical information to improve (lower) the scaling law (e.g. see Ierley \& Worthing (2001) for efforts in the Howard-Busse maximization problem). The best current bound on the Nusselt number - the ratio of actual heat flux to the conductive value -  is $Nu \leq 0.02634 Ra^{1/2}$ as $Ra \rightarrow \infty$ (PK03) whereas most of the current experimental data suggests $Nu \sim Ra^{0.31}$ (see the discussion in Waleffe et al. 2015)  and so is more consistent with the simple theoretical prediction of $Nu \sim Ra^{1/3}$  (Malkus 1954, Priestley 1954) with some dependence on the Prandtl number also possible (Grossmann \& Lohse 2000).  A natural way of incorporating further information exists in the background method through simply extending the definitions of the background fields. To see this, recall that the Malkus-Howard-Busse (maximization) approach and the Doering-Constantin (minimization) approach are dual problems seeking to find an  appropriate saddle point of a functional of the velocity and temperature fields (Kerswell 1998, 2001).  To explain further we introduce the problem to be considered.

Let a Newtonian fluid be confined between two infinite isothermal plates at $z=0$ and $z=d$ with the lower plate maintained at a constant temperature  $\delta T$ hotter than that of the upper plate (gravity is $-g \bhz$ where $g \approx 9.8ms^{-2}$). Using the gap width $d$, $d^2/\kappa$ ($\kappa$ is the thermal diffusivity) and $\delta T$ as units of length, time and temperature together with  adopting the Boussinesq approximation, the governing equations are
\beqa
(\N) &:=& \frac{\partial \bu}{\partial t}+\bu \cdot \bnab \bu+\bnab p-\sigma \nabla^2 \bu- \sigma Ra T \bhz={\bf 0},
\label{NS}\\
(\H) &:=& \frac{\partial T}{\partial t}+\bnab \cdot (\bu T-\bnab T)=0,
\label{Heat}
\eeqa
with $\bnab \cdot \bu=0$ where
\beq
\sigma:=\nu/\kappa \quad \& \quad  Ra:= g \beta \delta T d^3/ \nu \kappa
\eeq
are the Prandtl and Rayleigh numbers respectively ($\nu$ is the kinematic viscosity and  $\beta$ is the thermal expansion coefficient).  The background method starts by writing down the functional
\beq
\opL:=  \overline{\< |\bnab T|^2 \>}^t   -\overline{ \< a\bnu \cdot (\N)\> }^t - \overline{\<b\theta (\H)\>}^t
\eeq
where the first term on the right is the long-time-averaged Nusselt number $Nu$, $\bnu(\bx,t)$ and $\theta(\bx,t)$ are Lagrange multipliers imposing the  momentum and heat equations as constraints respectively (the seemingly redundant extra scalars $a$ and $b$ play a key role later) and the time and volume averages are defined as follows
\beq
\overline{(\,\, )}^t \, := \, \lim_{\T \rightarrow \infty} \sup \frac{1}{\T} \int^\T_0 (\, \, ) \,dt, \quad
\< \, \,\,\> \,  := \,\frac{1}{V} \int \, \, dV.
\eeq
The crucial next step is to define steady `background' fields
\beq
\bphi(\bx):=\bu(\bx,t)-\bnu(\bx,t), \quad \tau(\bx):=T(\bx,t)-\theta(\bx,t)
\eeq
which connect the Lagrange multipliers with the physical fields and such that they carry any inhomogeneous boundary conditions (so here just those on the temperature field).  Changing variables from $(\bu,T,\bnu,\theta)$ to $(\bu,T,\bphi,\tau)$,
\beqa
\opL                      &=& \overline{\< |\bnab T|^2 \>}^t  - \overline{\< a(\bu-\bphi) \cdot (\N)\>}^t - \overline{\< b(T-\tau) (\H)\>}^t , \nonumber \\
                              &=& \overline{\< |\bnab T|^2 \>}^t  - a\overline{\< \bu \cdot (\N)\>}^t +a \<\bphi \cdot \overline{(\N)}^t \>
                                                                              - b \overline{\<T(\H)\>}^t +b\<  \tau \overline{(\H)}^t \>
\eeqa
makes it clear that choosing the largest stationary value of  $\opL$ finds the largest long-time-averaged Nusselt number subject to the  long-time-averaged power and entropy balances (Lagrange multipliers $a$ and $b$ respectively) and projected information from momentum and heat flux balances (Lagrange multipliers $\bphi$ and $\tau$ respectively). Since it can be shown that all the time derivative terms in these constraints vanish under long time averaging,  the variational problem can be couched in terms of steady fields only. In particular, the goal is to evaluate the largest stationary value of the functional
\beq
 \opL_s:=\< |\bnab T|^2 \>  - a \< \bu \cdot (\N)_s\> +a \<\bphi \cdot (\N)_s \>
                                                                              - b \<T(\H)_s\> +b\<  \tau (\H)_s \>
\eeq
where the subscript $s$ indicates the steady version of the unsubscripted quantity. So far only the minimal choice $(\tau,\bphi)=(\tau(z),{\bf 0})$  has been  explored (Doering \& Constantin 1996) which leads to the simplified expression
\beq
\opL_s:=\< |\bnab T|^2 \>  - a \< \bu \cdot (\N)_s\> - b \<T(\H)_s\> +b\int^d_0 \tau(z) \overline{ (\H)_s}^{x,y} \, dz
\eeq
(where
\beq
\overline{(\,\, )}^{x,y} \, := \, \lim_{L \rightarrow \infty} \frac{1}{L^2} \int^{L/2}_{-L/2} \int^{L/2}_{-L/2}  (\, \, ) \,dx\,dy
\eeq
is a horizontal average). This choice turns out to give the dual problem to the Howard-Busse approach (Howard 1963, Busse 1969) and so produces the same Nusselt number bound (Kerswell 2001, PK03). However, here, beyond the total power and entropy balances and insisting that the fluid is incompressible and the boundary conditions are satisfied, only the horizontally-averaged heat equation is imposed as a constraint. It seems reasonable to suppose that imposing further constraints from  the governing equations by extending the definitions of the background fields should lower this current best bound for Boussinesq convection. Probing this hypothesis is the motivation for this paper.

This issue is quite general applying to bounds developed in other canonical flows such as plane Couette flow (Doering \& Constantin 1992), channel flow (Doering \& Constantin 1994) and pipe flow (Plasting \& Kerswell 2005). However, the most bounding work has been performed in  convection partly because of its wide application and partly because the current best bound appears to have the wrong exponent and therefore calls for the most improvement.  Of particular interest in recent efforts to lower the bound has been the introduction of the `wall-to-wall'  approach by Hassanzadeh et al. (2014) (see also Souza 2016 and Souza et al. 2019). Here the full heat equation has been imposed as a constraint with some incompressible boundary-compliant flow field which, apart from an overall amplitude,  is otherwise unconstrained and a {\em maximization} problem is solved. This appears to give a much improved upper bound of $Nu \sim Ra^{5/12}$ for stress-free boundary conditions in 2D convection with Souza (2016) finding a yet stronger (lower) bound  of $Nu \sim Ra^{0.371}$ for non slip boundary conditions. Later work by Tobasco \& Doering (2017), however, has  demonstrated through designing a sophisticated trial function that the upper bound must be at least $Nu \sim Ra^{1/2}$ up to logarithms for non-slip boundary conditions. This directly contradicts the conclusions of Souza (2016) and indirectly those of Hassanzadeh et al. (2014) (the heat flux for stress-free walls should be higher than for non-slip walls). Resolving this paradox by tackling the complementary background formulation - $(\tau,\bphi)=(\tau(x,z),{\bf 0})$ which imposes the full heat equation in 2D convection and builds a {\em minimization} problem - is a good starting point for our study.

Concurrent work by Souza et al. (2019) has considered how the background method is connected to the wall-to-wall approach and speculated that there could be a `duality gap' between them. Coming from a different perspective (the specific details of solving the variational equations), we share this speculation and confirm it here beyond a certain Rayleigh number. Motoki et al. (2018) have  also built upon  Hassanzadeh et al's work  by extending the maximization search to 3 dimensions. Interestingly they find a 3D optimal solution which scales like $Ra^{1/2}$ with a numerical coefficient just 7.2\% below the bound of PK03  (see their figure 2). This 3D result (using stress-free boundary conditions) and the 2D work of Tobasco \& Doering (2017) (using non-slip boundary conditions) clearly beg the question whether further information from the momentum equation can be used to rule out the $Ra^{1/2}$ scaling which clearly persists despite imposing the full heat equation. This also will be addressed here.

A further motivation for exploring the addition of further dynamical constraints is the hope that ultimately, the full steady governing equations can be imposed and then finally a direct connection forged between the optimal solution of an upper bounding variational problem and an actual solution of the (steady) governing equations.  On the one hand, it seems clear that the best bound possible using the background approach can't be lower than the highest heat flux attained by any of the many steady solutions of the governing equations whereas on the other, as argued above, the optimal solution should ultimately be a solution of the steady equations.
This possibility has been brought into  focus by recent computations tracking the (simple) 2D convection roll solution which initially bifurcates from the conductive state up to very high Rayleigh numbers of $O(10^9)$ (Waleffe et al. 2015, Sondak et al. 2015). Provided the aspect ratio of the rolls is optimized over, a heat flux relationship of $Nu \sim Ra^{0.31}$ is found which is intriguingly close to 3D turbulent convection measurements and $Nu \sim Ra^{1/3}$ which is  what many believe might be the ultimate scaling law although not all (e.g. Zhu et al. 2018).

A synopsis of the paper is as follows. Section 2 describes the set-up of 2D Boussinesq convection (\S2.1), explains how a bound can be found using the background approach (\S2.2) and then discusses the convexity of the optimization problem for a general temperature background field which ensures a unique optimal (\S2.3). Section 2.4 explains how the numerical computations are performed with a choice having to be made between a branch continuation approach  (PK03) and a time stepping method (Wen et al. 2013, 2015). Section 3 describes the results of tackling the upper bounding problem with the full heat equation imposed in the presence of the same symmetry as used in Hassanzadeh et al. (2014). The appearance of a second fluctuation mode becoming `spectrally unstable' at $Ra=Ra_c:=4468.8$ means: a) that Hassanzadeh et al's result is no longer a bound for $Ra> Ra_c$ (i.e. their result becomes only a  local maximum and the duality gap suggested by Souza et al. 2019 is realised); and b) a new formulation for how the optimal is tracked needs to be introduced compared to previous work (e.g. PK03). Section 4 discusses this new formulation which is significant because the various background and fluctuation optimal fields can no longer be used to define a set of physical temperature and velocity fields. In particular, the optimal fields do {\em not} satisfy the steady heat equation even though this is explicitly imposed as a constraint. Using this reformulation, section 5 shows how the optimal bound behaves for $Ra > Ra_c$. The size of the computational domain becomes important in the 2D background problem and it is found that the highest bound is only achieved in the infinite domain limit when the background field becomes increasingly 1D. Removing the symmetry used by Hassanzadeh et al. restores the translational invariance of the problem in which case the optimal has to be 1D and a bound of $Ra \le 0.055Ra^{1/2}$ is found compared to the well-known result of $0.026Ra^{1/2}$ for non-slip walls (PK03). Having found that imposing the full heat equation does not improve the bound, we then consider adding extra information from the momentum equation by introducing a background velocity field $\bphi(x,z)$. Now the optimization problem is no longer convex and so we are unable to invoke  uniqueness to dismiss non-vanishing $\bphi$. Instead we use an inductive bifurcation analysis to show that if $\bphi={\bf 0}$ before a bifurcation then it remains ${\bf 0}$ after it too meaning that the continuous branch of optimals found by branch tracking out of the energy stability point always has $\bphi={\bf 0}$. Noting the one caveat that it's not  impossible that there is an unconnected branch of optimals with $\bphi \neq {\bf 0}$, this strongly suggests the surprising result that imposing the full Boussinesq equations does not improve the bound over that obtained using the horizontally-averaged Boussinesq equations.
Finally, section 7 observes that adding a velocity background temperature field to the formulation of Wen et al. (2015), which has an additional vorticity constraint, also fails to improve matters. A discussion follows in section 8.

%
%
\section{Mathematical formulation}
%


\subsection{Set-up}

We consider the 2 dimensional version of the Boussinesq equations (\ref{NS}) \& (\ref{Heat}) where $\boldsymbol{u}=u\boldsymbol{\hat{x}}+w\boldsymbol{\hat{z}}$ over a box $(x,z) \in [-\half L, \half L] \times [0,1]$ together with the following stress-free boundary conditions
\beqa
\frac{\partial u}{\partial z}=w=0, &&\quad T=1,\quad at\quad z=0, \label{4}\\
\frac{\partial u}{\partial z}=w=0,&&\quad T=0,\quad at\quad z=1 \label{5}
\eeqa
following Hassanzadeh et al. (2014). Applying the background method, we decompose the temperature field as
\begin{equation}\label{8}
  T=\tau(x,z)+\theta(x,z,t).
\end{equation}
where the (steady) background temperature $\tau$ carries the boundary conditions of $T$ (i.e. $\tau|_{z=0}=1$ and $\tau|_{z=1}=0$) so that the perturbation field $\theta$  vanishes at $z=0,1$. The time-averaged heat transport is characterized by the time-averaged Nusselt number $Nu$
\begin{equation}\label{10}
 Nu:=\lim_{T\rightarrow\infty}\frac{1}{T}\int^T_0 \frac{1}{L}\int^{L/2}_{-L/2} \,\frac{\partial T}{\partial z}
\biggl|_{z=1} \biggr.\, dx\,dt=\langle|\boldsymbol{\nabla}T|^2\rangle=1+\langle wT\rangle
\end{equation}
in which henceforth
\begin{equation}\label{10b}
\nonumber
   \langle \,(\ldots)\, \rangle :=\lim_{T\rightarrow\infty}\frac{1}{T}\int^T_0\int^1_0 \frac{1}{L}\int^{L/2}_{-L/2} \,(\ldots)\, dx \, dz\, dt
\end{equation}
is the spatial-temporal average. To find the maximum  heat transport possible over all solutions to the Boussinesq equations, we construct the Lagrangian
\beqa
 \opL && =\langle|\boldsymbol{\nabla}T|^2\rangle-\frac{a}{\sigma Ra} \langle \boldsymbol{u}\cdot \N \rangle-b\langle \theta \,\H \rangle,\\
              && = \langle|\boldsymbol{\nabla}T|^2\rangle-\frac{a}{\sigma Ra} \langle \boldsymbol{u}\cdot \N \rangle-b\langle T\,\H \rangle +b \langle  \tau\, \H\rangle. \label{11}
\eeqa
where  $a/\sigma Ra$ is the Langrange multiplier imposing the global constraint $\langle \boldsymbol{u}\cdot \N \rangle=0$, $b$ is a Lagrange multiplier imposing the global constraint $\langle T\,\H \rangle=0$ and $b \tau(x,z)$ is the Lagrange multiplier field imposing the time-averaged heat equation pointwise in the domain. The inclusion of $b$ is actually redundant given the constraint imposed by $\tau$ implies $\langle T\,\H \rangle=0$ so the value of $b$ is chosen for convenience. Expression (\ref{11}) can be rewritten using integration by parts and the fact that $\langle wT\rangle={\cal L}-1$ (see (\ref{10})) for solutions of the Boussinesq equations as
\begin{equation}\label{13}
   \opL= \frac{1}{1-a}\left[\langle{|\boldsymbol{\nabla}\tau|}^2\rangle-a \right]-\frac{1}{1-a}\mathscr{G}
\end{equation}
where  setting $b=2$ makes
\begin{equation}\label{14}
  \mathscr{G}:=\langle \frac{a}{Ra}|\boldsymbol{\nabla}\boldsymbol{u}|^2+|\boldsymbol{\nabla}\theta|^2+2\theta \boldsymbol{u}\cdot\boldsymbol{\nabla}\tau \rangle
\end{equation}
a purely quadratic form in $\theta$ and $\boldsymbol{u}$.

\subsection{Bounds on $Nu$}

The key realisation is that {\em if} $\mathscr{G} \ge 0$ for all
$(\boldsymbol{u}, \theta) \in \Pi$ (the set of incompressible velocity and temperature fields which satisfy homogeneous versions of the boundary conditions (\ref{4}) and (\ref{5})\,), which is a {\it spectral constraint} on $\tau$ and $a \in (0,1)$, we then have the bound
\beq
Nu \, \leq \,  \frac{1}{1-a}\left[\langle{|\boldsymbol{\nabla}\tau|}^2\rangle-a \right].
\eeq
The challenge is then to find the lowest such bound by minimizing over all $(\tau,a)$ which satisfy this spectral constraint, i.e.
\begin{equation}
(\tau,a) \in \Omega:=\{ (\tau,a) \,\,|\,\, \mathscr{G}(\boldsymbol{u},\theta;\tau,a) \geq 0 \, \, \forall \, (\boldsymbol{u},\, \theta) \in \Pi\,\}.
\end{equation}
After introducing a streamfunction, $(u,w)=(\partial\psi/\partial z,-\partial\psi/\partial x$), the constraint that
\begin{equation}
\mathscr{G}=  \langle \frac{a}{Ra}|\nabla^2 \psi|^2+|\boldsymbol{\nabla}\theta|^2+2\theta J(\tau,\psi) \rangle  \geq 0
\end{equation}
where
\beq
J(A,B):=\frac{\partial A}{\partial x}\frac{\partial B}{\partial z}-\frac{\partial A}{\partial z}\frac{\partial B}{\partial x}
\eeq
is equivalent to requiring that all of the  eigenvalues $\lambda$ of
the linear problem
\beqa
 \lambda\theta && =\nabla^2\theta-J(\tau, \psi), \label{15d}\\
  \lambda\nabla^2\psi && =\frac{a}{Ra}\nabla^4\psi-J(\tau,\theta) \label{15e}
\eeqa
(with boundary conditions $\psi=d^2 \psi/dz^2=\theta=0$ for $z=\{0,1\}$) are negative semi-definite.

\subsection{Convexity \& Uniqueness}

The Euler-Lagrange equations for stationarizing  the Lagrangian $\opL$ in (\ref{13}) are
\beqa
0 &&= \nabla^2\theta-J(\tau,\psi),                                                                                              \label{EL_1}\\
0 &&= \frac{a}{Ra}\nabla^4 \psi-J(\tau,\theta),                                                                         \label{EL_2}\\
0 &&= \nabla^2\tau -J(\theta, \psi),                                                                                            \label{EL_3}\\
\langle{ |\boldsymbol{\nabla}\tau|}^2\rangle-1  &&= \frac{(1-a)}{Ra}\langle |\nabla^2 \psi|^2 \rangle   \label{EL_4}
\eeqa
and, as a nonlinear set of equations, can have many solutions. However, only solutions with $(\tau,a) \in \Omega$ yield a bound through the value of $\opL$ generated. Due to the  convexity of $\Omega$ (i.e. if  $(\tau_1,a_1)$ and $(\tau_2,a_2)$ are in $\Omega$  then so is $\lambda (\tau_1,a_1)+(1-\lambda) (\tau_2,a_2)$  for $\lambda \in (0,1)$), and the fact that the objective functional
\begin{equation}
f(\tau,a):=\frac{1}{1-a}\left[\langle{|\boldsymbol{\nabla}\tau|}^2\rangle-a \right]
\label{objective}
\end{equation}
to be minimized is a strictly convex functional (the terms second order in $\delta \tau$ and $\delta a$ in the difference $f(\tau+\delta \tau, a+\delta a)-f(\tau,a)$, specifically
\beq
\frac{1}{(1-a)^2} \langle\, \boldsymbol{\nabla} | (1-a)\delta \tau +( \tau+z-1) \delta a \,|^2 \,\rangle,
\eeq
are positive definite), there is in fact at most one solution which satisfies the spectral constraint. This solution, hereafter referred to as the {\em optimal solution}, is what is sought.

\subsection{Numerical approach}

Recently, \cite{Wen} have proved that when $\tau$ is 1-dimensional, i.e. $\tau=\tau(z)$, appropriately augmenting the (steady) Euler-Lagrange equations with time derivatives leads to a system where  the optimal solution is a unique attracting steady state.
This proof carries over to 2-dimensional background fields $\tau=\tau(x,z)$ in the 3-dimensional Rayleigh-Benard problem but not in the 2-dimensional problem (see appendix A for details) where the dimensionality of the background field then matches that of the physical fields. This means that  any steady attractor which emerges from time-stepping using
$\tau(x,z)$ in the 2D problem is not guaranteed to be the required optimal solution. The time stepping approach can still be used if it is married with a spectral constraint check but then there is always the  prospect of rerunning with different initial conditions until the optimal solution is found. Given this, we chose instead to use the branch continuation approach - Newton's method with parametric continuation - starting from the energy stability bifurcation point as performed in PK03.  While very robust, this has the general disadvantage of only being able to continuously trace optimal solutions from the energy stability bifurcation as $Ra$ varies meaning that  any new unconnected optimals cannot be found at a given $Ra$. This is not a problem here as  the aforementioned uniqueness of the optimal solution means that no other optimal solution branches exist.

We consider periodic boundary conditions in $x$ and, exactly as in Hassanzadeh et al. (2014), assume that the streamfunction $\psi$ is odd (or antisymmetric), while $\theta$ and $\tau$ are even (or symmetric) about $x=0$ by seeking the solution of (\ref{EL_1})-(\ref{EL_4}) in the following form:
\begin{equation}\label{21}
\psi=\sum^M_{m=0} \psi_m(z)         \sin(m\alpha x),\quad
\theta=\sum^M_{m=0} \theta_m(z) \cos(m\alpha x),\quad
\tau=\sum^M_{m=0} \tau_m(z)        \cos(m\alpha x).
\end{equation}
We will find that this choice prevents a 1D background optimal  even  though this is allowed by the boundary conditions and imposed symmetry.
Here $\alpha:=2\pi/L$ and $\psi_m$, $\theta_m$, $\tau_m$ are expanded in Chebyshev polynomials, $T_n$,
\beq
\label{22}
[\psi_m(z),\theta_m(z),\tau_m(z)]=\sum^N_{n=0} [\psi_{mn},\theta_{mn},\tau_{mn}]T_n(z)
\eeq
where $T_n(z):=\cos(n\cos^{-1}(2z-1))$. Resolution varies from $(N,M)=(30,30)$ to $(80,80)$ to ensure numerical accuracy as $Ra$ increases and $L$ changes .

%
%
\begin{figure}
 \centering
  \scalebox{0.8}[0.8]{\includegraphics[bb=0 0 413 448]{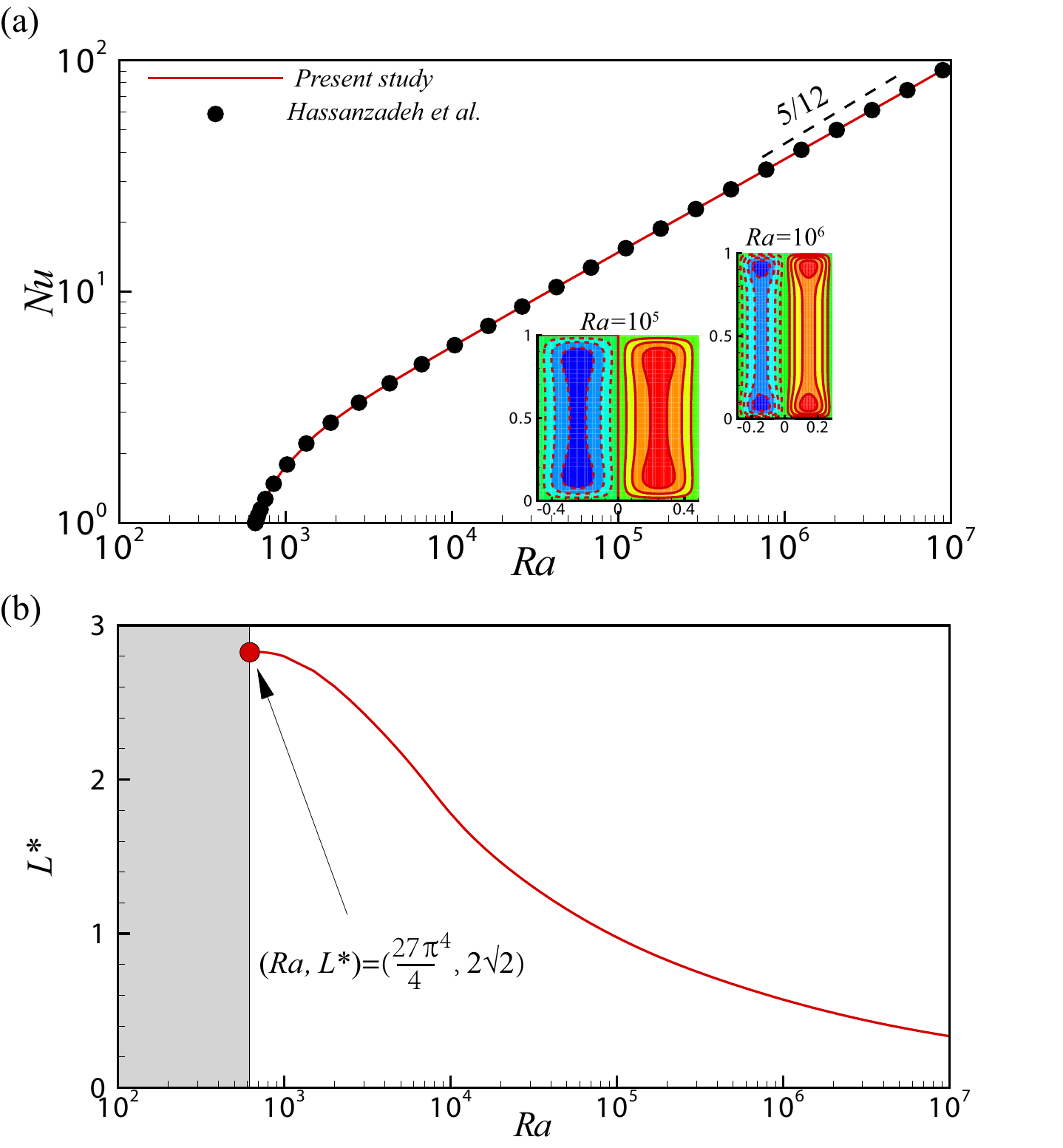}}
 \caption{\label{fig02} (a) The Nusselt number vs. the Rayleigh number tracked up from the energy stability bifurcation point. The solution is optimized over the domain length and the data is in excellent agreement with Hassanzadeh \etal (2014). Subplots show the flow streamfunctions at $Ra=10^5$ and $10^6$ (data courtesy of Dr A. Souza). (b) The optimal domain size $L*$ vs. the Rayleigh number. The bullet is the energy stability bifurcation point.}
\end{figure}
%
%
\begin{figure}
 \centering
  \scalebox{0.65}[0.65]{\includegraphics[bb=0 0 413 215]{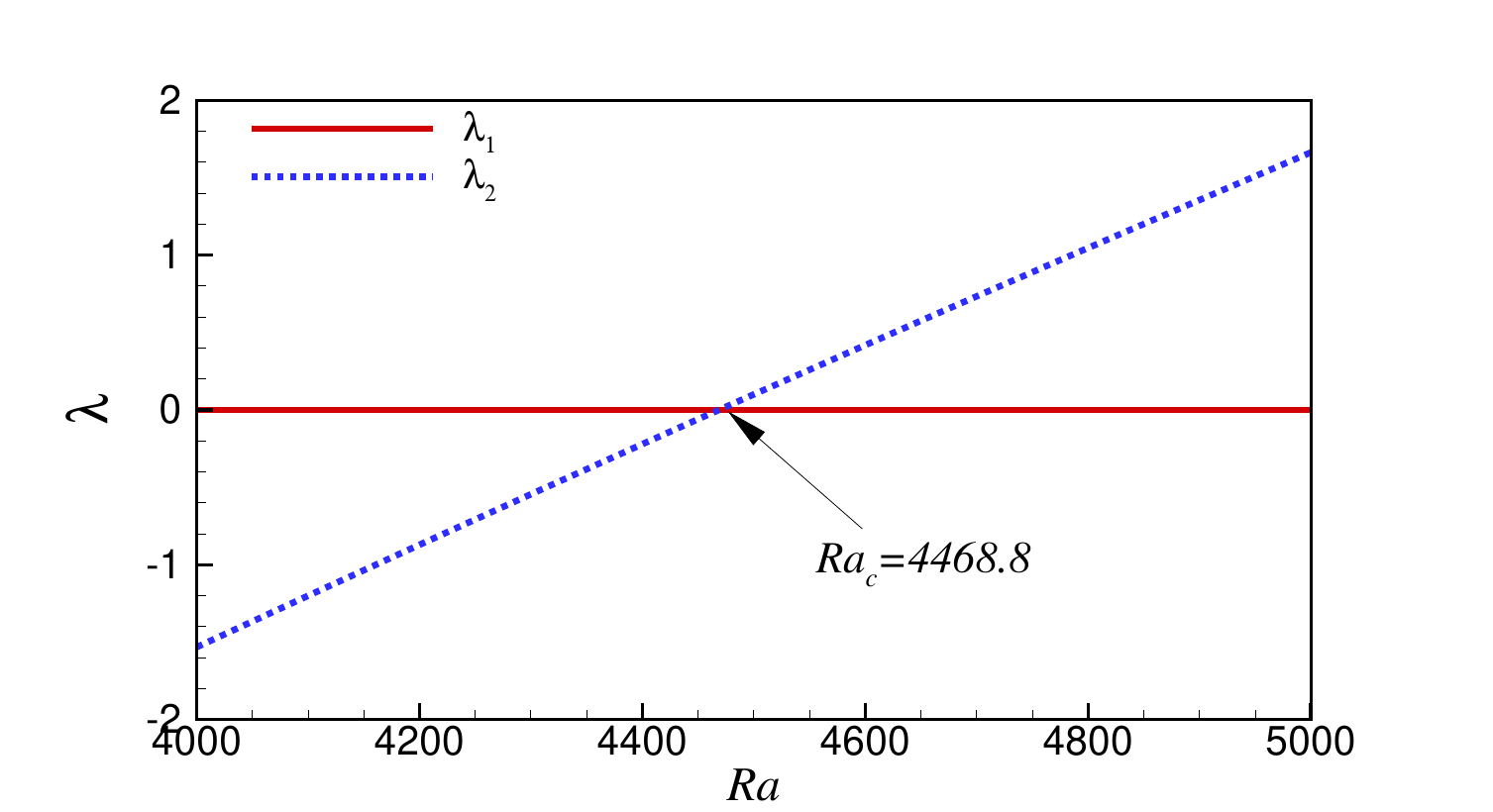}}
\caption{\label{fig03} The first ($\lambda_1$) and second ($\lambda_2$) largest eigenvalues of the spectral constraint for $Ra < 4468.8$. At $Ra=4468.8$, where they cross, the aspect ratio is $L^*=2.234$.}
\end{figure}

%
%
\section{Connecting to Hassanzadeh \etal (2014) }\label{sec3}
%

The conductive temperature profile $\tau=1-z$ is a spectrally-stable background field until  $Ra=27\pi^4/4$ in a domain of  size $L=2\sqrt{2}$ when energy instability first starts. Pinning the marginal fluctuation fields $(\theta,\psi)$ - hereafter a {\em mode} - as was done in PK03, the optimal solution was then tracked up to $Ra=10^7$ with the domain size $L=L^*(Ra)$ simultaneously optimized to yield the highest heat flux at a given  $Ra$: see figure \ref{fig02}.  The calculated $Nu$ values correspond exactly with those found by Hassanzadeh et al. (2014) (as do flow fields computed at $Ra=10^5$ and $10^6$; see the inset of \ref{fig02}). This indicates that Hassanzadeh et al.'s (2014) wall-to-wall transport approach is equivalent to the background method when a single mode is considered.

%
%
\begin{figure}
 \centering
  \scalebox{0.95}[0.95]{\includegraphics[bb=0 0 400 193]{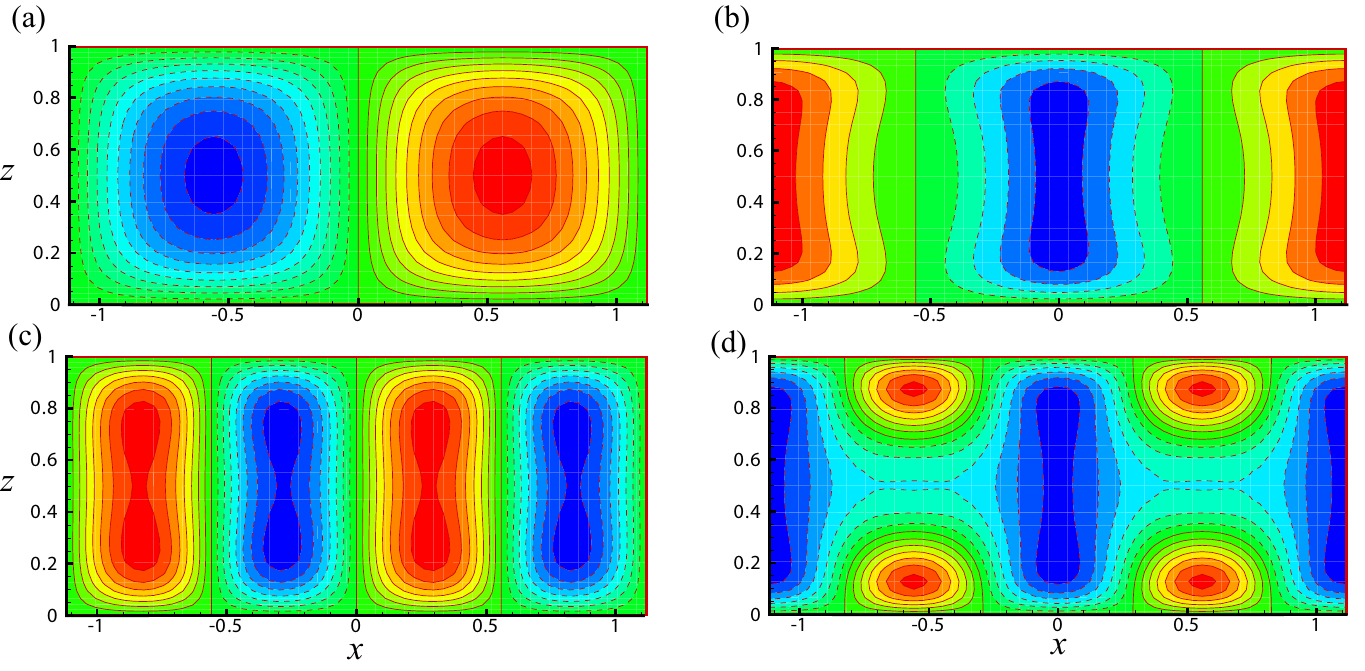}}
 \caption{\label{fig04} At $Ra=4468.8$, the first mode (a) $\psi_1$ and (b) $\theta_1$ and the new second mode (c) $\psi_2$ and
 (d) $\theta_2$.}
\end{figure}

\begin{figure}
 \centering
 \scalebox{0.65}[0.65]{\includegraphics[bb=75 0 464 243]{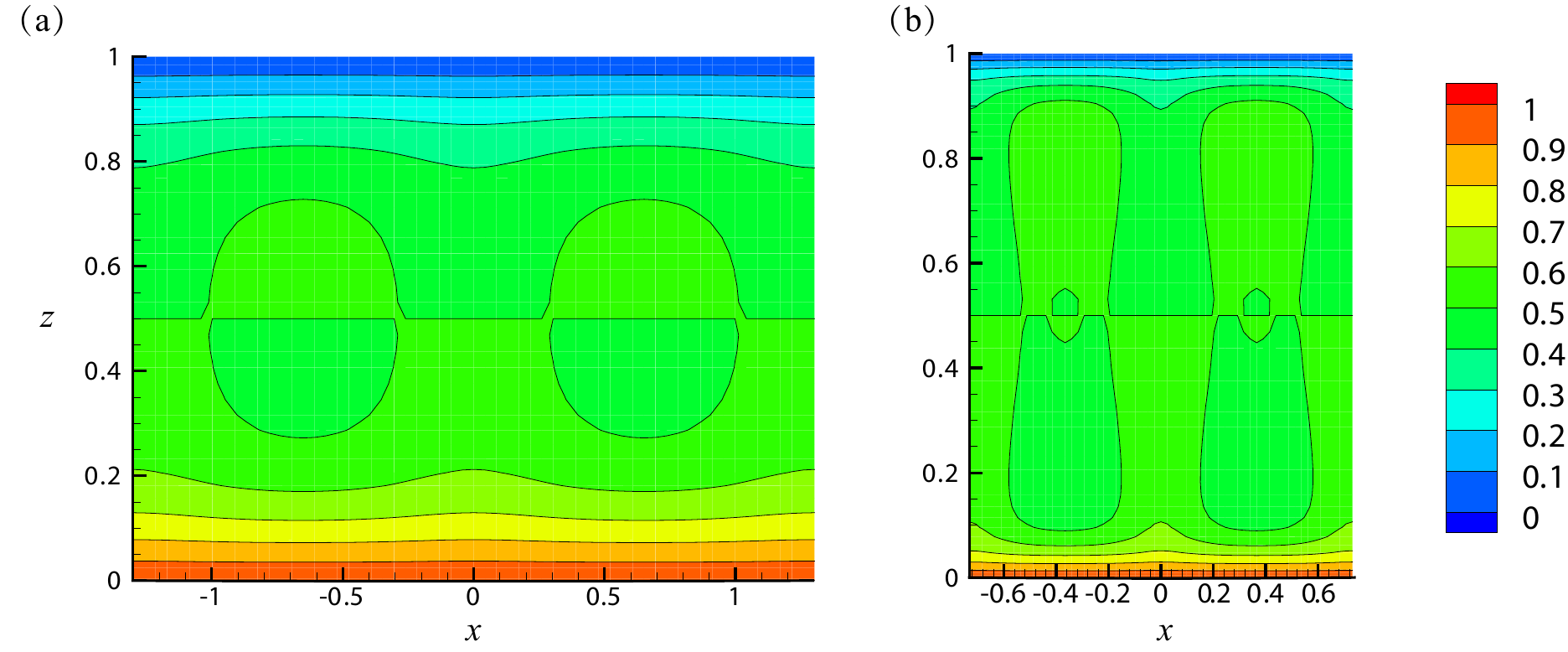}}
 \caption{\label{figtau} The optimal background field $\tau$ plotted at (a) $Ra=2000$ and (b) $Ra=20000$.}
\end{figure}

In their wall-to-wall optimal control approach, however, Hassanzadeh et al. (2014) had no way of identifying whether their local optimal was in fact the global optimal. It should be sufficiently close to the energy stability point but experience in other related problems (e.g. PK03) suggests that further modes in  the spectral constraint eventually become marginal as $Ra$ increases. The optimal solution should subsequently modify itself to keep  these new modes marginal with concomitant adjustments in the $Nu$-scaling. Fortunately, in the background approach, the spectral constraint provides a check on whether a given Euler-Lagrange solution is the optimal solution. Solving the eigenvalue problem (\ref{15d})-(\ref{15e}) for disturbances which are also periodic over $[0,L^*(Ra)]$ demonstrates that the eigenvalue ($\lambda_1$ in figure \ref{fig03}) of the first mode is pinned at 0, while a second mode becomes marginal at $Ra=4468.8$ for an aspect ratio $L^*=2.234$.  This means that \cite{Hassanzadeh}'s result is \emph{not} a bound for $Ra>4468.8$. This apparently is not because any further 2D bifurcation has been missed (Chini, {\em private communication}) but more a reflection of the `duality gap' suggested by Souza et al. (2019) being realised. Figure \ref{fig04}(a,b) shows the first mode $(\psi_1,\theta_1)$ with wavenumber $\alpha_1:=2\pi/L^*$ so the flow field contains one pair of convection cells. The second mode ($\psi_2,\theta_2$) with $\alpha_2=2\alpha_1$ illustrated in figure \ref{fig04}(c,d) has two pairs of convection cells.  The optimal background field at $Ra=2000$ is shown in figure \ref{figtau} and for the now non-optimal 1-mode solution at $Ra=20,000$. In both cases the field is weakly 2-dimensional indicating that the first mode consists of   non-monochromatic (i.e. non-single $\alpha$) velocity and temperature fields. The emergence of the second mode at $Ra=4468.8$ indicates that the background profile is now degenerate  in a way which has important implications for solving the Euler-Lagrange equations for higher $Ra$ while respecting the spectral constraint. We discuss this in the following section.

%
%

%
%
\section{Multi-modal optimals}\label{rub}
%
%

When a new mode becomes marginal  in the spectral constraint as the background field $\tau$ evolves with $Ra$, a further `pinning' constraint needs to be added to keep the new mode marginal in the spectral constraint as $Ra$ increases further. This procedure is thoroughly discussed in \cite{Doering4} and implemented in PK03 for a background field of lower dimensionality than the fluctuation field. In this situation, an example of which is using a 1D case $\tau=\tau(z)$ in the 2D Rayleigh-Benard problem, the fluctuation field can be Fourier-transformed over the spatial dimension(s) across which $\tau$ is invariant and then considered parameterized by the Fourier wavenumber $k$. Different spectrally-marginal fluctuation fields  have different $k$ and are then naturally orthogonal under averaging over this spatial dimension. This means that  the Euler Lagrange equations (\ref{EL_1})-(\ref{EL_4}),
\beqa
0 &&= \nabla^2\theta_j-J(\tau,\psi_j),                                                      \qquad j=1,\ldots, N                              \label{EL_1_1D}\\
0 &&= \frac{a}{Ra}\nabla^4 \psi_i-J(\tau,\theta_i),                                 \qquad j=1,\ldots, N                              \label{EL_2_1D}\\
0 &&= \tau_{zz} -\sum_{j=1}^N\overline{J(\theta_j, \psi_j)},                                                                                  \label{EL_3_1D}\\
\langle |\tau_z |^2 \rangle-1  &&= \frac{(1-a)}{Ra} \sum_{j=1}^N \langle |\nabla^2 \psi_j|^2 \rangle    \label{EL_4_1D},
\eeqa
(the overbar represents averaging over $x$) can simply be extended to include the new marginal mode
\beq
(\theta_{N+1},\psi_{N+1})(x,z)=(\hat{\theta}_{N+1}, \hat{\psi}_{N+1})(z)e^{ik_{N+1}x}
\label{monochromatic}
\eeq
when it appears. Equivalently, the Lagrangian is just
\begin{equation}
\opL = \frac{1}{1-a} \left[ \langle {\tau_z}^2 \rangle-a \right]-\frac{1}{1-a}\mathscr{G}
\end{equation}
where  $\mathscr{G}$ naturally partitions into the contributions from the various marginal modes as follows
\beq
 \mathscr{G}=  \langle \frac{a}{Ra}|\nabla^2 \psi|^2+|\boldsymbol{\nabla}\theta|^2+2\theta J(\tau,\psi) \rangle
= \sum_{j=1}^{N+1}  \mathscr{G}_j
\eeq
where
\beq
\mathscr{G}_j:=\langle \frac{a}{Ra}|\nabla^2 \psi_j|^2+|\boldsymbol{\nabla}\theta_j|^2+2\theta J(\tau,\psi_j) \rangle.
\eeq
The appearance of a new spectrally-marginal mode merely extends the set  of wavenumbers contributing  to the definition of the fluctuation field by one,
\beq
(\psi, \theta) (x,z):= \sum_{j=1}^{N+1} (\psi_j,\theta_j)=\sum_{j=1}^{N+1} (\hat{\psi}_j,\hat{\theta}_j)(z) e^{i k_j x}.
\eeq
Importantly, this means it is possible to talk about the unique optimal solution of the variational problem which satisfies the imposed physical constraints as  being
\beq
(\psi,T)(x,z)=(0,\tau)(z)+\sum_{j=1}^N (\hat{\psi}_j,\hat{\theta}_j)(z) e^{i k_j x},
\eeq
i.e. the spectral constraint is satisfied at a saddle point of $\opL$.

This pleasing situation in which the marginal fluctuation fields have a physical interpretation changes, however,  when the dimensionality of the background field equals the dimensionality of the problem (the case here), or, pathologically, there is more than one marginal mode for a {\em given} wavenumber (see chapter 3 of Fantuzzi 2018).  In these  scenarios, the natural orthogonality property of different marginal fluctuation fields disappears with the result that  the physical meaning of the fluctuation fields is lost. To see this, the key is to realise that pinning the marginal fluctuation fields is done (Doering \& Constantin 1996) as before by writing the Lagrangian as
\begin{equation}\label{25}
\mathscr{L}=\frac{1}{1-a}(\langle{|\boldsymbol{\nabla}\tau|}^2\rangle-a)-\frac{1}{1-a}\sum_{j=1} \mathscr{G}_j.
\end{equation}
The constraint that each $\mathscr{G}_j$ vanishes pins the $j^{th}$ mode to be marginal (the Lagrange multiplier imposing this is absorbed into the amplitude of the $j^{th}$ marginal fluctuation field) while $\mathscr{G} > 0 $ for all other fluctuation fields.
However, since the modes $(\psi_j,\theta_j)$ are not now orthogonal,
\begin{equation}
\sum_{j=1}^{N} \mathscr{G}_j  \neq \mathscr{G} :=  \langle \frac{b}{Ra}|\nabla^2 \psi|^2+|\boldsymbol{\nabla}\theta|^2+2\theta J(\tau,\psi) \rangle
\label{key}
\end{equation}
for $N\geqslant 2$ where
\beq
(\psi,\theta)(x,z)=\sum_{j=1}^N (\psi_j,\theta_j)(x,z)
\eeq
is taken as the total optimal fluctuation field. In fact, $\mathscr{G} > 0$ and so this total optimal field is {\em not} a solution of the Heat  equation. The clear implication is that the spectral constraint is not satisfied for $N \geqslant 2$ at {\em any} saddle point of the Lagrangian (\ref{13})  where the steady heat  equation is  imposed. Consequently, the optimization procedure is forced to find an optimal away from the  saddle points of the Lagrangian (\ref{13}) where the spectral constraint is satisfied to deliver a bound.

From a different perspective, Souza et al. (2019)  have also recently argued that this should happen when exploring the connection between the wall-to-wall approach (a max-min problem) with the associated background method (a min-max problem). A duality gap means that
\beq
({\rm wall-}{\rm to}{\rm-wall}) \qquad \sup_{\theta,\bu} \, \inf_{\tau} \mathscr{L} \,<\,  \inf_{\tau} \, \sup_{\theta,\bu} \mathscr{L} \qquad ({\rm background})
\eeq
(making the connection $\eta=\tau-(1-z)$ and $\zeta=\theta$ with the variables used by Souza et al 2019) where the optimal solution to the wall-to-wall problem {\em is} achieved at a stationary point of $\mathscr{L}$ thereby implying  that to the background method is not.
They also supply a nice simple quadratic polynomial in 5 variables to illustrate the phenomenon.  The calculations described in the next section confirm that this gap starts to exist as soon as $N =2$.

\section{Extending Hassanzedah {\em et al.} with a symmetric 2D background field $\tau(x,z)$}   \label{odc1}

\begin{figure}
 \centering
  \scalebox{0.75}[0.75]{\includegraphics[bb=0 0 508 194]{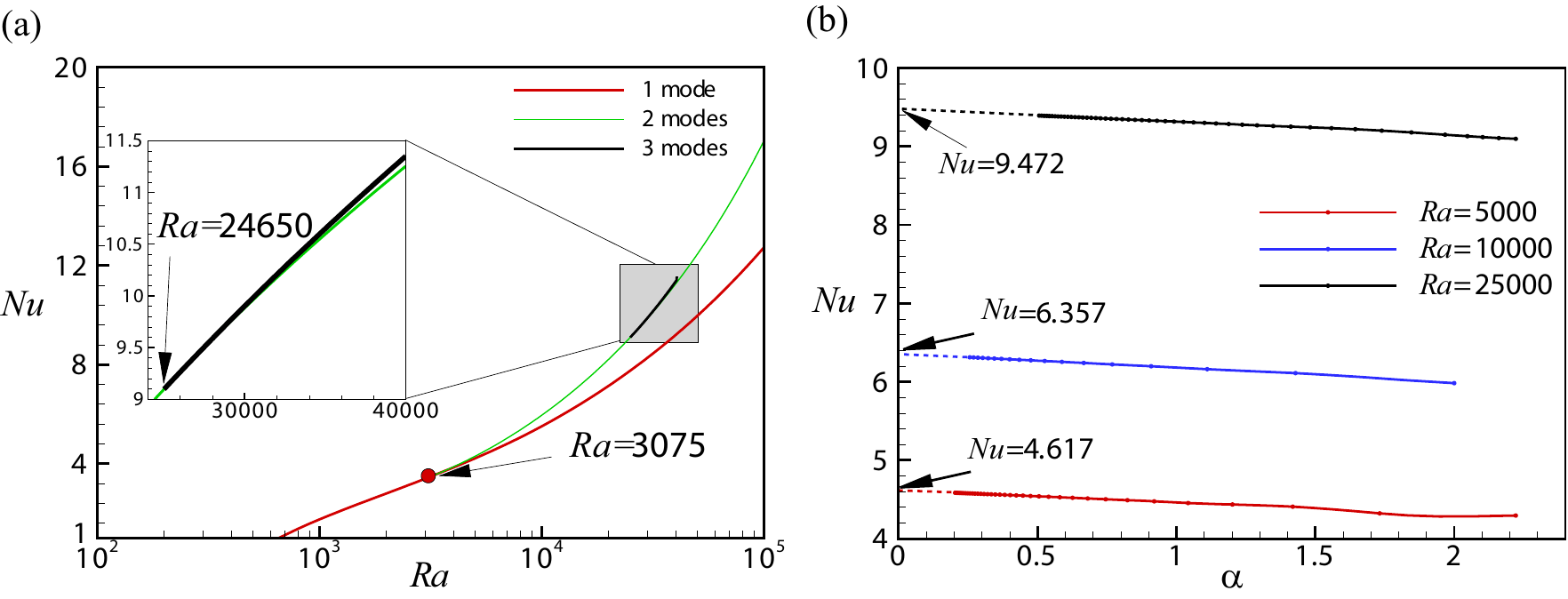}}
 \caption{\label{fig05} (a) The Nusselt number  $Nu$ vs. the Rayleigh number $Ra$ at fixed aspect ratio $L=2\sqrt{2}$. (b) $Nu$ vs. $\alpha=2\pi/L$.  }
\end{figure}

To explore multi-modal bounding solutions, a first series of computations were done in the fixed domain  $L=2 \sqrt{2}$. In this geometry, the first mode appears  at $Ra=27 \pi^4/4$ (the energy stability threshold), the second mode at $Ra=3,075$ and the third mode at $Ra=24,650$. The 1-mode and 2-mode optimal solution branches could be easily continued up to $Ra=10^5$ whereas the 3-mode solution branch proved difficult to continue much beyond $Ra>40,000$ due to numerical issues: see figure \ref{fig05}(a). The 3-mode solution, which provides an upper bound in this geometry over at least the range $24,650 \leq Ra \leq 40,000$, presents only a modest correction to the 2-mode optimal solution which is no longer a bound for these $Ra$.

A second series of computations were  then carried out to investigate the dependence of the $Nu$-bound on the aspect ratio $L$.  Three different $Ra$ values were chosen to explore  the dependence of the bound on $L$: $Ra=5000$ and $10,000$ where the bound is given by a 2-mode solution, and $Ra=25,000$ where the bound is given by a 3-mode solution.
In all three cases, the largest bound is  achieved as  the aspect ratio $L \rightarrow \infty$: see figure \ref{fig05}(b). This is very different from the optimal control results of \cite{Hassanzadeh} where the optimal  aspect ratio scales like $ Ra^{-1/4}$ and so vanishes  as $Ra \rightarrow \infty$.

\begin{figure}
 \centering
  \scalebox{0.7}[0.7]{\includegraphics[bb=0 0 544 138]{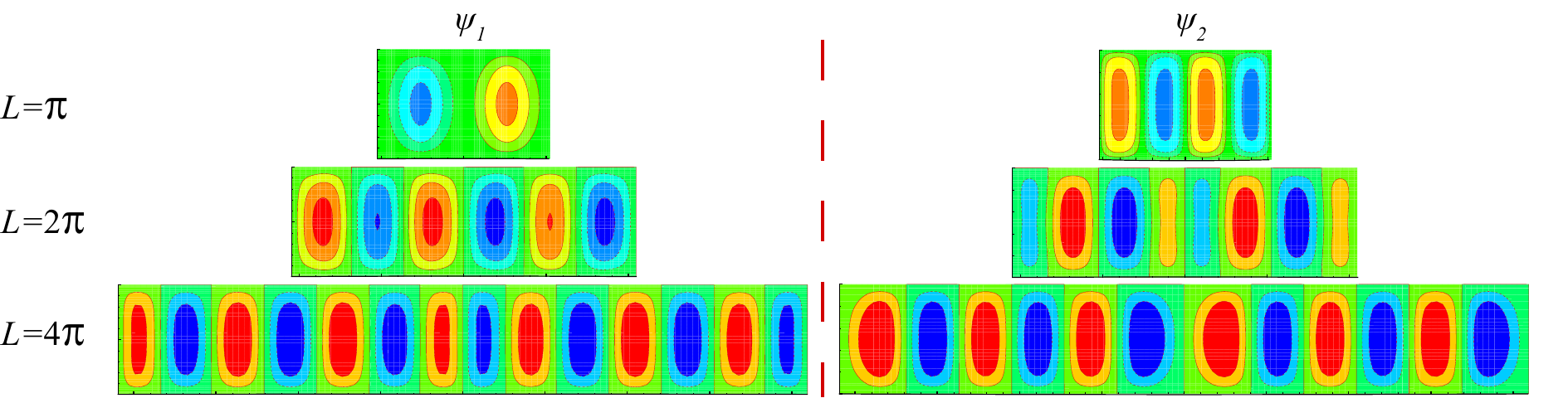}}
   \scalebox{0.7}[0.7]{\includegraphics[bb=0 0 544 138]{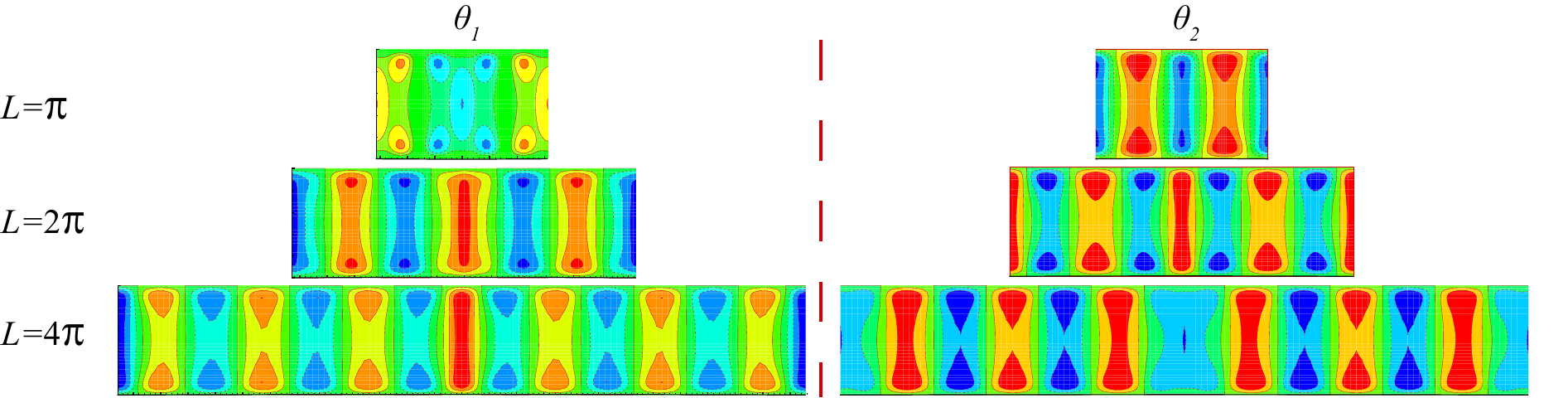}}
 \caption{\label{fig06} Left panel: The first mode ($\psi_1,\theta_1$); right panel: the second mode ($\psi_2,\theta_2$) at $Ra=10^4$ (only two critical modes are present for this $Ra$).}
\end{figure}

\begin{figure}
 \centering
  \scalebox{0.75}[0.75]{\includegraphics[bb=0 0 492 181]{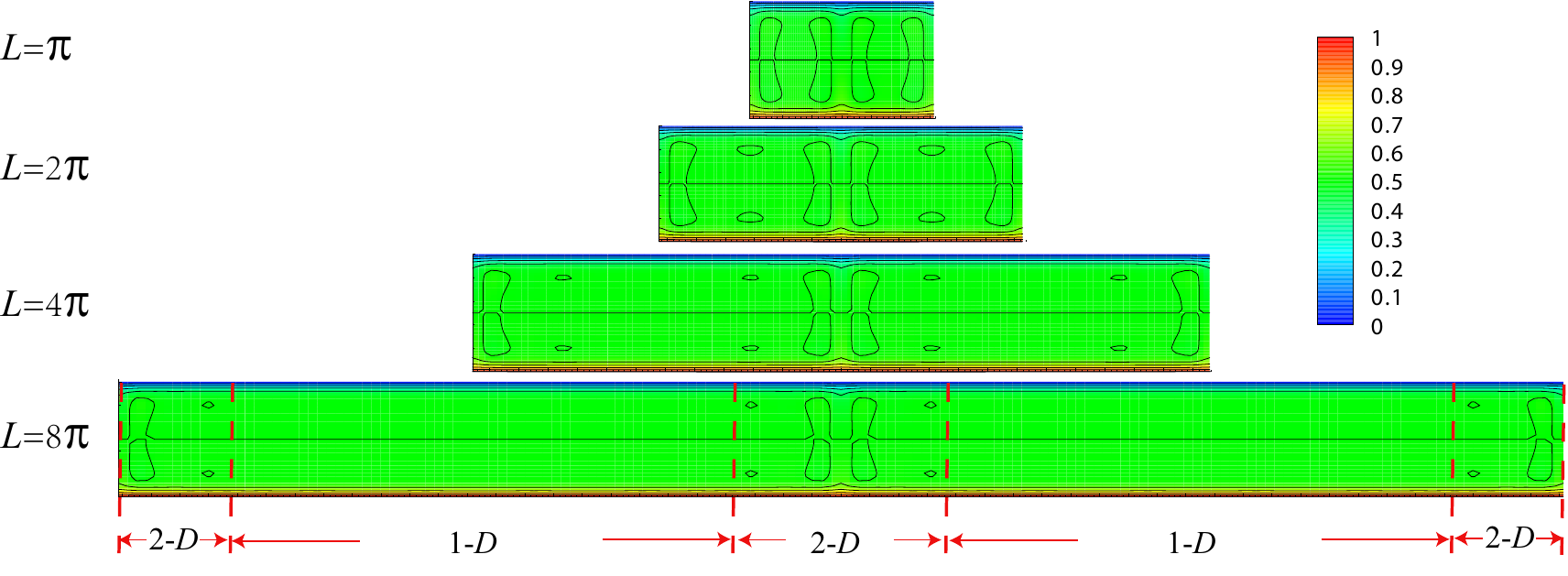}}
 \caption{\label{fig07} The background field plotted at different aspect ratios at $Ra=10^4$.}
\end{figure}

\begin{figure}
 \centering
  \scalebox{0.75}[0.75]{\includegraphics[bb=0 0 479 129]{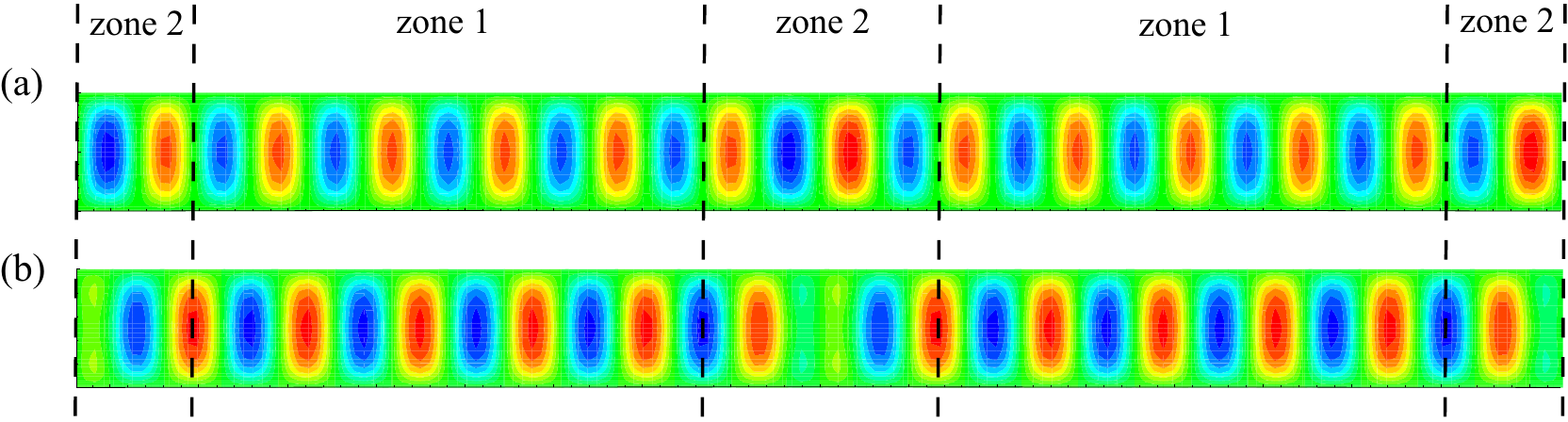}}
 \caption{\label{fig08} The two critical flow fields at $Ra=10^4$ and $L=8\pi$: (a) $\psi_1$ and (b) $\psi_2$.}
\end{figure}

Figure \ref{fig06} shows the  structure of the two modes at $Ra=10^4$. The fluctuation fields $\psi_i$ and $\theta_i$  for both $i=1$ and $2$ have  a convection roll structure and  increasing $L$ just means that more of the rolls fit into the domain. On closer inspection it is clear that the rolls are slightly different near to  $x=0$ and $x=\pm\frac{1}{2}L$) where they are forced to have a certain symmetry (symmetry around $x=0$ and periodicity over a length $L$ force symmetry about $x=\pm \frac{1}{2}L$ as well).  When the domain is short, e.g. $L=\pi$, the background field is clearly two-dimensional as seen in figure \ref{fig07}. However, as $L$ increases to $L=8\pi$, the background field become predominantly one-dimensional ($1$-$D$) away from the imposed lines of symmetry at $x=0$ and  $x=\pm\frac{1}{2}L$ (the ends of the domain shown). Plotting the streamfunctions $\psi_1$ and $\psi_2$ over this long domain - see figure \ref{fig08} - confirms that the convection cells are similar  away from the symmetry lines (`zone 1' in figure \ref{fig08}) where $\tau$ is predominantly $1$-$D$ but are quite different close to the symmetry lines (`zone 2') where $\tau$ is clearly $2$-$D$.

The  structure of the optimal fields (both background and fluctuation) and the fact that the bound is maximised as $L \rightarrow \infty$ indicate that the optimal solution is trying to minimise the effect of the imposed symmetry requirements at $x=0$ and $x=\pm \frac{1}{2}L$. Without this imposed symmetry, the problem becomes translationally invariant  and the optimal solution must be $1$-$D$ by the convexity result in \S 2.3. There is another  simple way to see this. Since the bounding functional $f(\tau,a)$ (see (\ref{objective})\,) is strictly convex in both $\tau(x,z)$ and $a$, any  $2$-$D$ solution $(\tau_{2D}(x,z),a) \in \Omega$
\begin{equation}
f \biggl( \frac{1}{N} \sum_{j=1}^{N} \tau_{2D}(x+\tfrac{jL}{N},z), a \biggr)
\, < \,
\frac{1}{N} \sum_{j=1}^N f \biggl( \tau_{2D}(x+\tfrac{jL}{N},z),a \biggr)
\end{equation}
by Jensen's inequality. Taking the limit of $N \rightarrow \infty$ in the left hand side and using translational invariance of the problem in the right hand side leads to
\begin{equation}
f \biggl( \tau_{1D}(z):=\frac{1}{L} \int_{0}^{L} \tau_{2D}(x,z)\, dx, a \biggr) \,<\, f(\tau_{2D}(x,z),a)
\end{equation}
so that a $1$-$D$ background field  always produces a better bound than a $2$-$D$ field. The results in figure \ref{fig07} indicate that this is what the optimal solution is trying to achieve.

%
%

%
%

%
%

%
%

\subsection{Lifting the symmetry: 1D background field}

\begin{figure}
 \centering
  \scalebox{0.85}[0.85]{\includegraphics[bb=0 0 218 201]{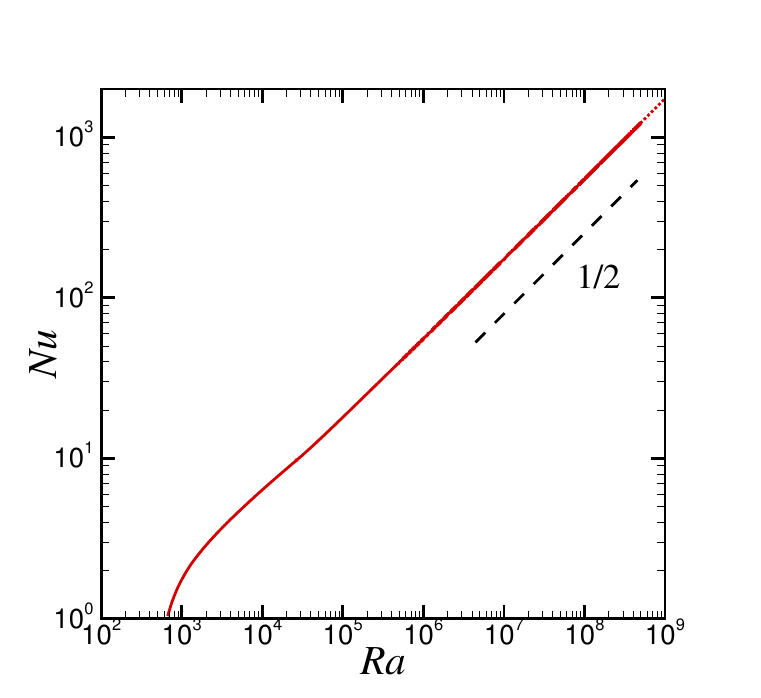}}
   \scalebox{0.85}[0.85]{\includegraphics[bb=0 0 218 201]{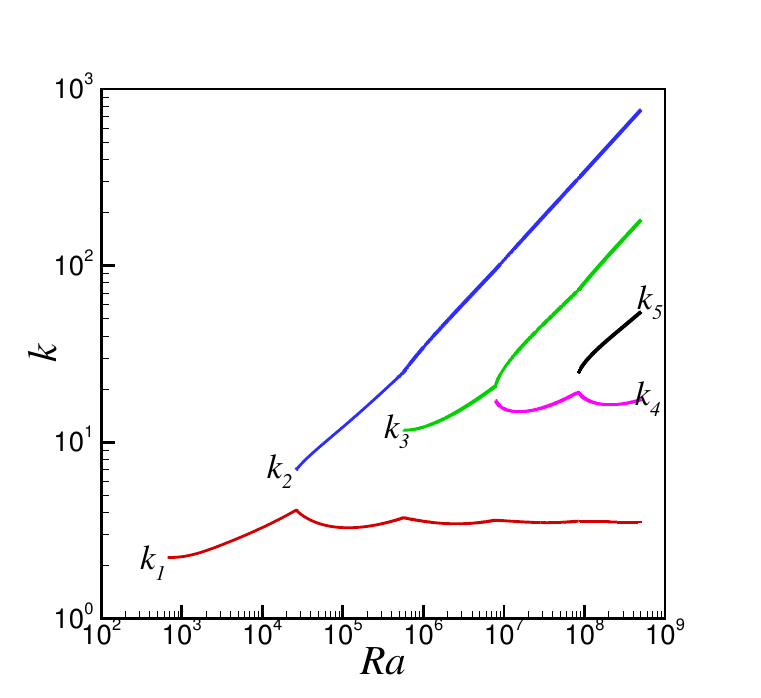}}
 \caption{\label{fig10} Left panel: the upper bound of $Nu$ vs. the Rayleigh number $Ra$ with $Nu\le 0.055Ra^{1/2}$ in the asymptotic regime. right panel: the bifurcation diagram of critical wave number $k_m$ vs. the Rayleigh number. }
\end{figure}

Calculation of the optimal solution assuming from the onset that the background field is $1$-$D$ simplifies the computation since the fluctuation fields can then be parametrised by their single wavenumber in $x$ (as in (\ref{monochromatic})). In this case, rather than setting a domain periodicity and insisting the fluctuation wavenumbers be consistent with this, the wavenumbers themselves can be optimised over as real continuous variables meaning, in effect, that $L$ is infinite. For example, the Euler-Lagrange equation corresponding to the $m^{th}$ wavenumber $k_m$ is
\begin{equation}\label{39}
  \delta \mathscr{L}/\delta k_m:=-2\int^1_0 ak_m(u_m^2+w_m^2)+k_m\theta_m^2dz+\int^1_0 p_m u_m dz=0.
\end{equation}
With this formulation, Newton's method with branch continuation proved much faster than the time-stepping approach. It took  around $4$ hours cputime on a 2.6Ghz laptop using Newton's method to obtain the optimal solution from $Ra=27\pi^4/4$ up to $Ra=5\times10^8$ while the time-stepping approach took  at least a day to generate a single point at $5 \times Ra=10^8$. However, when the domain is fixed, Newton's method becomes very inefficient as the critical wavenumbers $k_m$ are discrete and cannot be tracked using a (continuous) continuation method: in this case, time-stepping is the better choice.  The numerical solution of the one-dimensional background problem gives the upper bound of $Nu \le 0.055Ra^{1/2}$ as shown in figure \ref{fig10}(a) with 5 critical modes present by $Ra=10^9$ (\,see figure \ref{fig10}(b)\,). This result has the same scaling exponent as the non-slip result $Nu \leq 0.026 Ra^{1/2}$ of PK03 but with a  larger numerical coefficent as should be expected for stress-free boundary conditions. The prior work of \cite{Wen} indicates that adding a further enstrophy constraint (possible only in stress-free 2D convection) significantly improves the bound obtained here down to $Nu \leq 0.106 Ra^{5/12}$.
%

%
%

It is worth briefly discussing how the critical wavenumbers which appear in the 1-D and 2-D background field calculations are related. Figure \ref{fig10}(b) indicates that at  $Ra=10^4$ there is only one critical wavenumber $k_1=3.284$ for the 1-D background problem.  However, for the 2-D (symmetric) background problem, there are two critical modes as seen in figure \ref{fig06} and figure \ref{fig08} with both having an approximate wavenumber $\approx 3.3$. Both these modes are forced to be antisymmetric (and so in phase) about $x=0$ and $x=\pm L/2$ (zone 2 in figure \ref{fig08})
but away from these points endeavour to be approximately $\pi/2$ out of phase (zone 1 in figure \ref{fig08}). With this phase difference together with matching amplitudes so
\begin{align}
\psi_1 & =f(z)\sin(k_1x),\qquad \qquad\theta_1=g(z)\cos(k_1x), \label{42a}\\
\psi_2 &=f(z)\sin(k_1x+\pi/2),\quad \theta_2 =g(z)\cos(k_1x+\pi/2), \label{42b}
\end{align}
the nonlinear term in Eq.(\ref{EL_3})
\begin{equation}\label{43}
\sum^2_{i=1}
\biggl[ \frac{\partial\psi_i}{\partial z}\frac{\partial\theta_i}{\partial x}-\frac{\partial\psi_i}{\partial x}\frac{\partial\theta_i}{\partial z} \biggr]=-k_1 \biggl[ \frac{df}{dz}g+\frac{dg}{dz}f \biggr]
\end{equation}
is  $x$ independent and therefore can only drive a 1-D background field. From another perspective, the 1-D background field problem really has two modes with $k=3.284$ but only one needs to be tracked as the nonlinear term is horizontally averaged (e.g. see (\ref{EL_3_1D})\,) ensuring that the background field stays 1-D.

\begin{figure}
 \centering
  \scalebox{0.75}[0.75]{\includegraphics[bb=0 0 459 334]{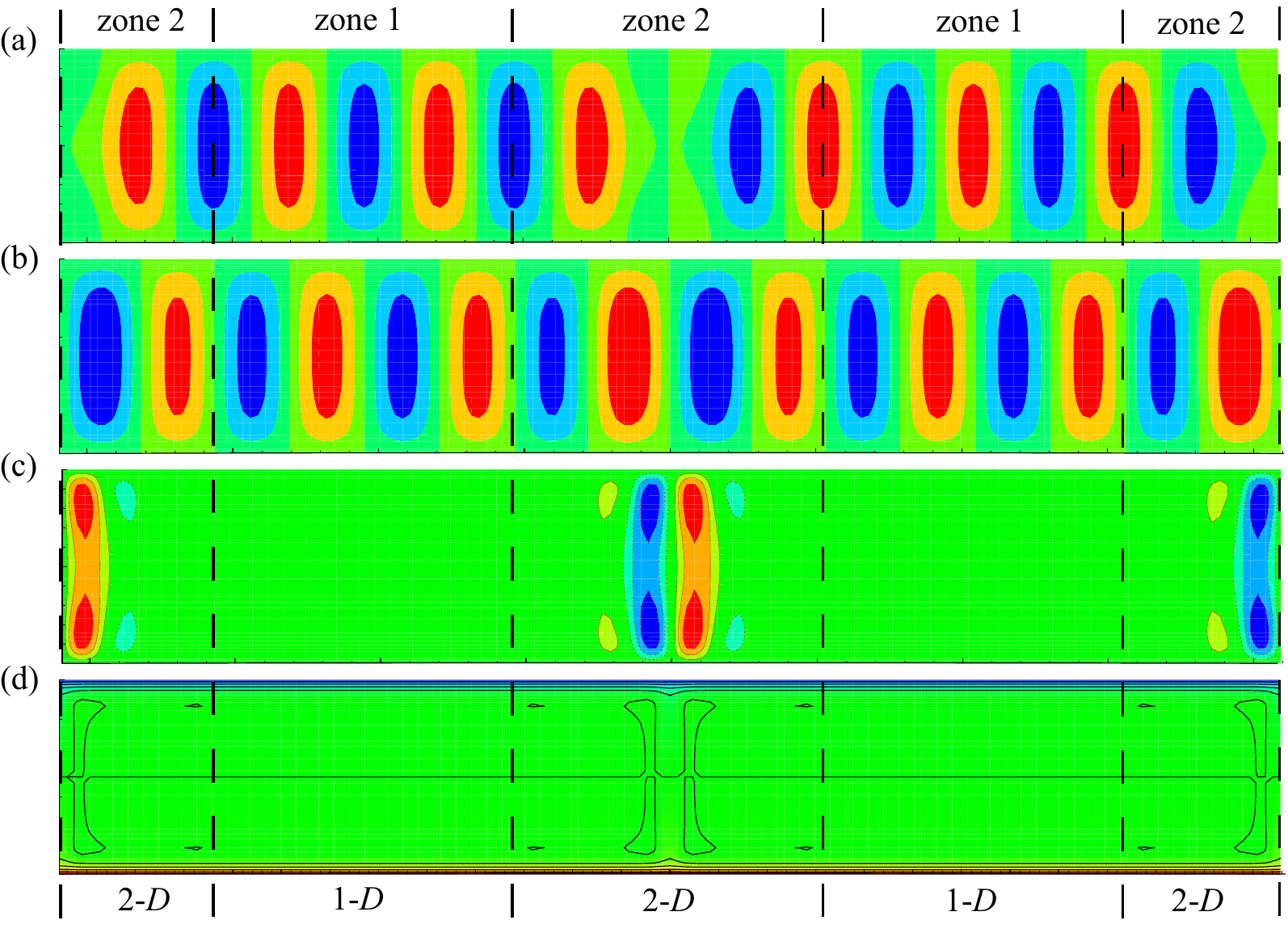}}
 \caption{\label{fig11} (a-c) The profiles of the three critical flow fields $\psi_1$, $\psi_2$, $\psi_3$ at $Ra=25000$, $L=4\pi$. (d) The profile of the two-dimensional optimal background field.}
\end{figure}

At  $Ra=25,000$ there is even a third mode in the 2-D background problem compared to  still only 1 mode in the 1-D background problem (the second wavenumber $k_2$ appears at $Ra \approx 26,450$). Figure \ref{fig11} shows that in fact $\psi_3$  is only significant in zone 2 where the imposed symmetries dominate. In zone 1  where the background field is essentially 1-D profile, $\psi_3$ vanishes.

The conclusion of the computations so far is that imposing the full heat equation in the bounding calculation does {\em not} improve (lower) the bound over that obtained using the horizontally-averaged heat equation. The next obvious question is whether this is also true when imposing  the full momentum equation. The next section addresses this.

%
%

\section{Imposing the full momentum equation: $\bphi \neq {\bf 0}$}

In this section, we attempt to improve the bound by using a background temperature field {\em and} a background velocity field of the same dimension as the physical problem which means  that the full momentum equation and heat equation are imposed as constraints.
Importantly, the optimization problem is no longer convex and so we are unable to invoke  uniqueness to dismiss $\bphi$. Instead we use an inductive bifurcation analysis to show that if $\bphi={\bf 0}$ before a bifurcation then it remains ${\bf 0}$ after it too meaning that the continuous branch of optimals found by branch tracking out of the energy stability point always has $\bphi={\bf 0}$.

The analysis begins by constructing the following Lagrangian:
\begin{equation}\label{two2}
\mathscr{L}=\langle|\boldsymbol{\nabla}T|^2\rangle
-\frac{a}{\sigma Ra}\langle\boldsymbol{v}\cdot \boldsymbol{\mathcal{N}}\rangle-2\langle\theta \,\mathcal {H}\rangle
\end{equation}
which, after introducing the extended background decomposition
\begin{equation}\label{two1}
\boldsymbol{u} = \boldsymbol{\phi}+\boldsymbol{v},\quad T=\tau+\theta,
\end{equation}
can be  rewritten as
\begin{equation}
\mathscr{L}=\frac{\langle|\boldsymbol{\nabla}\tau|^2\rangle-a}{1-a}-\frac{a}{1-a}\langle\phi_3\tau\rangle-\frac{1}{1-a}\mathscr{G}
\end{equation}
where
\begin{multline}
\nonumber
\mathscr{G}(\boldsymbol{v},\theta):=\langle 2\theta(\boldsymbol{\phi}+\boldsymbol{v})\cdot\boldsymbol{\nabla}\tau+|\boldsymbol{\nabla}\theta|^2\rangle+a\langle\phi_3\theta\rangle\\+\langle\frac{a}{\sigma Ra}\boldsymbol{v}\cdot\boldsymbol{v}\cdot\boldsymbol{\nabla}\boldsymbol{\phi}+\frac{a}{\sigma Ra}\boldsymbol{v}\cdot\boldsymbol{\phi}\cdot\boldsymbol{\nabla}\boldsymbol{\phi}+\frac{a}{ Ra}|\boldsymbol{\nabla}\boldsymbol{v}|^2-\frac{a}{Ra}\boldsymbol{v}\cdot\nabla^2\boldsymbol{\phi}\rangle
\end{multline}
(note $\mathscr{G}$ depends parametrically on $\tau$, $\boldsymbol{\phi}$, $a$, $\sigma$ and $Ra$ but this is suppressed for clarity). If  $\inf_{\boldsymbol{v},\theta} \mathscr{G}$ exists (and necessarily $0<a<1$), a bound is then given by
\begin{equation}
Nu\le\frac{\langle|\boldsymbol{\nabla}\tau|^2\rangle-a}{1-a}-\frac{a}{1-a}\langle\phi_3\tau\rangle-\frac{1}{1-a}\inf_{\boldsymbol{v}, \theta} \mathscr{G}(\tau,\boldsymbol{\phi}).
\end{equation}
The difficulty here is that the objective functional is no longer convex and so it's unclear how to establish a priori that the optimal solution takes the form  $(\tau,\bphi)=(\tau(z),{\bf 0})$.  Hence we consider what has to happen at bifurcation points.
Minimization of $\mathscr{G}$ with respect to incompressible $\boldsymbol{v}$ and $\theta$ requires
\begin{equation}\label{two5a}
-{\frac{2a}{Ra}}\nabla^2\boldsymbol{v}+\frac{a}{\sigma Ra}\boldsymbol{v}\cdot(\boldsymbol{\nabla}\boldsymbol{\phi}+\boldsymbol{\nabla}\boldsymbol{\phi}^\texttt{T})
+\frac{a}{\sigma Ra}\boldsymbol{\phi}\cdot\boldsymbol{\nabla}\boldsymbol{\phi}-\frac{a}{\sigma Ra}\nabla^2\boldsymbol{\phi}+2\theta\boldsymbol{\nabla}\tau+\boldsymbol{\nabla}p=0,
\end{equation}
\begin{equation}\label{two5b}
\boldsymbol{\nabla}\cdot\boldsymbol{v}=0,
\end{equation}
\begin{equation}\label{two5c}
-2\nabla^2\theta+a\phi_3+2(\boldsymbol{v}+\boldsymbol{\phi})\cdot\boldsymbol{\nabla}\tau=0
\end{equation}
the solution of which is denoted as $(\boldsymbol{v}_0,\theta_0)$.
The Lagrangian can then be written as
\begin{equation}\label{twolang}
\mathscr{L}=\frac{\langle|\boldsymbol{\nabla}\tau|^2\rangle-a}{1-a}-\frac{a}{1-a}\langle\phi_3\tau\rangle-
\frac{1}{1-a} \biggl\{ \mathscr{G}(\boldsymbol{v}_0,\theta_0; \tau, \boldsymbol{\phi})
+\sum_{i=1}^N \mathscr{H}(\boldsymbol{v}_i,\theta_i) \biggr\}
\end{equation}
where $\boldsymbol{v}_0$, $\boldsymbol{v}_i$ ($i=1\ldots N$) and $\boldsymbol{\phi}$ are incompressible fields and
%
%
\begin{equation}
\mathscr{H}(\boldsymbol{v},\theta):=\langle 2\theta \boldsymbol{v} \cdot\boldsymbol{\nabla}\tau+|\boldsymbol{\nabla}\theta|^2+\frac{a}{\sigma Ra}\boldsymbol{v} \cdot\boldsymbol{v} \cdot\boldsymbol{\nabla}\boldsymbol{\phi}+\frac{a}{ Ra}|\boldsymbol{\nabla}\boldsymbol{v}|^2 \rangle
\end{equation}
is a purely quadratic functional of $(\boldsymbol{v},\theta)$ which must be positive semi-definite - the spectral constraint - for $\inf \mathscr{G}$ to exist. The  fields $(\boldsymbol{v}_i,\theta_i)$ are marginal in that $\mathscr{H}(\boldsymbol{v}_i,\theta_i)=0$ and their number $N$ increases with $Ra$. The aim is to minimize the upper bound over $\tau$, $\boldsymbol{\phi}$ and  $a$ at fixed $\sigma$ and $Ra$ subject to this spectral constraint.
The Euler-Lagrange equations are:
the spectral constraint equations for $\boldsymbol{v}_i$ and $\theta_i$
\begin{equation}\label{apt1}
-(1-a)Ra \frac{\delta \mathscr{L}}{\delta\boldsymbol{v}_i}:=-2a\nabla^2\boldsymbol{v}_i+\frac{a}{\sigma}\boldsymbol{v}_i\cdot(\boldsymbol{\nabla}\boldsymbol{\phi}+\boldsymbol{\nabla}\boldsymbol{\phi}^\texttt{T})
+2Ra\theta_i \boldsymbol{\nabla}\tau+\boldsymbol{\nabla}p_i ={\bf 0}, \qquad i=1 \ldots N
\end{equation}
\begin{equation}\label{apt2}
-(1-a)\frac{\delta \mathscr{L}}{\delta\theta_i}:=-2\nabla^2\theta_i+2\boldsymbol{v}_i\cdot\boldsymbol{\nabla}\tau=0.  \qquad i=1 \ldots N;
\end{equation}
the forced field equations for $\boldsymbol{v}_0$ and $\theta_0$
\begin{equation}\label{apt3}
-(1-a)Ra\frac{\delta \mathscr{L}}{\delta\boldsymbol{v}_0}:=-2a\nabla^2\boldsymbol{v}_0+\frac{a}{\sigma}\boldsymbol{v}_0\cdot(\boldsymbol{\nabla}\boldsymbol{\phi}+\boldsymbol{\nabla}\boldsymbol{\phi}^\texttt{T})
+2Ra\theta_0\boldsymbol{\nabla}\tau+\boldsymbol{\nabla}p_0\\+\frac{a}{\sigma }\boldsymbol{\phi}\cdot\boldsymbol{\nabla}\boldsymbol{\phi}-a\nabla^2\boldsymbol{\phi}={\bf 0};
\end{equation}
\begin{equation}\label{apt4}
-(1-a)\frac{\delta \mathscr{L}}{\delta\theta_0}:=-2\nabla^2\theta_0+2\boldsymbol{v}_0\cdot\boldsymbol{\nabla}\tau+2\boldsymbol{\phi}\cdot\boldsymbol{\nabla}\tau+a\phi_3=0;
\end{equation}
the background field equations
\begin{multline}\label{apt5}
(1-a)Ra\frac{\delta \mathscr{L}}{\delta\boldsymbol{\phi}}:=a\nabla^2\boldsymbol{v}_0-aRa(\tau+\theta_0)\boldsymbol{e}_z-2Ra\theta_0\boldsymbol{\nabla}\tau+\boldsymbol{\nabla}q+\frac{a}{\sigma }\boldsymbol{v}_0\cdot\boldsymbol{\nabla}\boldsymbol{v}_0\\-\frac{a}{\sigma }(\boldsymbol{v}_0\cdot\boldsymbol{\nabla}\boldsymbol{\phi}^\texttt{T}-\boldsymbol{\phi}\cdot\boldsymbol{\nabla}\boldsymbol{v}_0)+\frac{a}{\sigma }\sum_{i=1}^N\boldsymbol{v}_i\cdot\boldsymbol{\nabla}\boldsymbol{v}_i={\bf 0},
\end{multline}
\begin{equation}\label{apt6}
(1-a)\frac{\delta \mathscr{L}}{\delta \tau}:=-2\nabla^2\tau-a\phi_3+2(\boldsymbol{v}_0+\boldsymbol{\phi})\cdot\boldsymbol{\nabla}\theta_0+2\sum_{i=1}^N\boldsymbol{v}_i\cdot\boldsymbol{\nabla}\theta_i=0;
\end{equation}
and finally the balance parameter equation
\begin{multline}\label{apt6b}
(1-a)\frac{\delta \mathscr{L}}{\delta a}:= \mathscr{L}-1 -\langle\phi_3\tau\rangle
-\biggl\{
\sum_{i=1}\langle\frac{1}{\sigma Ra}\boldsymbol{v}_i\cdot\boldsymbol{v}_i\cdot\boldsymbol{\nabla}\boldsymbol{\phi}+\frac{1}{Ra}|\boldsymbol{\nabla}\boldsymbol{v}_i|^2 \rangle \\
+\langle\phi_3\theta_0+\frac{1}{Ra}|\boldsymbol{\nabla}\boldsymbol{v}_0|^2+\frac{1}{\sigma Ra}\boldsymbol{v}_0\cdot\boldsymbol{v}_0\cdot\boldsymbol{\nabla}\boldsymbol{\phi}+\frac{1}{\sigma Ra}\boldsymbol{v}_0\cdot\boldsymbol{\phi}\cdot\boldsymbol{\nabla}\boldsymbol{\phi}-\frac{1}{Ra}\boldsymbol{v}_0\cdot\nabla^2\boldsymbol{\phi} \rangle
\biggr\}.
\end{multline}
The pressure-like quantities have been rescaled $Ra\,p_i\rightarrow p_i$, $Ra \, p_0 \rightarrow p_0$ and $(1-a)Ra \,q\rightarrow q$) and incompressibility conditions on $\boldsymbol{v}_0$, $\boldsymbol{v}_i$ and $\boldsymbol{\phi}$ are left implicit. A key point here is that the forced field pair $(\boldsymbol{v}_0,\theta_0)$ is {\it not} marginal in the spectral constraint i.e. $\mathscr{H}(\boldsymbol{v}_0,\theta_0) >0$.

\subsection{The first bifurcation point}\label{fbp}

The solution at the first critical point $Ra_c=27\pi^4/4$ is $\tau=1-z$ and $\boldsymbol{\phi}=0$, $(\boldsymbol{v}_0,\theta_0)=({\bf 0},0)$ and $a=1$. At $Ra=Ra_c$, the spectral constraint becomes marginal for the first time, i.e. there is a non-trivial solution to the spectral problem
\begin{equation}\label{apt7}
-2\nabla^2\boldsymbol{v}_i+2Ra\theta_i\boldsymbol{\nabla}\tau+\boldsymbol{\nabla}p_i=0,
\end{equation}
\begin{equation}\label{apt8}
\boldsymbol{\nabla}\cdot\boldsymbol{v}_i=0,
\end{equation}
\begin{equation}\label{apt9}
-\nabla^2\theta_i+\boldsymbol{v}_i\cdot\boldsymbol{\nabla}\tau=0.
\end{equation}
There are two different modes (using symmetries):
\begin{equation}\label{apt10}
(\boldsymbol{v}_1, \theta_1)=A_1(\, U(z)\sin(kx)\boldsymbol{e}_x+W(z)\cos(kx)\boldsymbol{e}_z,\Theta(z)\cos(kx)\,)
\end{equation}
\begin{equation}\label{apt11}
(\boldsymbol{v}_2, \theta_2)=A_2(\, U(z)\cos(kx)\boldsymbol{e}_x-W(z)\sin(kx)\boldsymbol{e}_z,-\Theta(z)\sin(kx)\,)
\end{equation}
Since $d\tau/dz=-1$, the structure in $z$ is simple: $U:=\pi\cos(\pi z)$, $W:=-k\sin(\pi z)$ and  $\Theta:=-\frac{\sqrt{2}}{3\pi}\sin(\pi z)$ where $k=\pi/\sqrt{2}$.
Slightly away from the critical point, $Ra=Ra_c+\eps$, the fields need to be  expanded as follows
%
%
\begin{align}
\tau                         &=\tau_0+\eps \tau_1+\eps^2 \tau_2 +\dots,  \nonumber \\
\boldsymbol{\phi} &=\qquad \eps \boldsymbol{\phi}_1+\eps^2 \boldsymbol{\phi}_2+\eps^3 \boldsymbol{\phi}_3\dots, \nonumber \\
\boldsymbol{v}_0  &=\qquad  \eps \boldsymbol{v}_0^1+\eps^2 \boldsymbol{v}_0^2+\dots, \nonumber \\
\theta_0                 &= \qquad \eps \theta_0^1                 +\eps^2 \theta_0^2+\dots, \nonumber \\
\boldsymbol{v}_i  &= \eps^{1/2} \boldsymbol{v}_i^0+\eps^{3/2} \boldsymbol{v}_i^1+\eps^{5/2} \boldsymbol{v}_i^2+\dots, \nonumber \\
\theta_i                   &=\eps^{1/2} \theta_i^0+\eps^{3/2} \theta_i^1+\eps^{5/2} \theta_i^2+\dots, \nonumber\\
a                             &=\;a_0+\eps  a_1 +\eps^2 a_2+\dots \nonumber
\end{align}
where $\tau_0:=1-z$ and $a_0:=1$.

\subsubsection{Leading Order}

To leading order, the spectral constraint is satisfied by $(\boldsymbol{v}^0_i, \theta^0_i)$ defined in (\ref{apt10}-\ref{apt11}) and these fields force the other leading order equations for the background fields
%
%
\begin{align}
2\nabla^2\tau_1+\phi_{13} &=2\sum_{i=1,2}\boldsymbol{v}_i^0\cdot\boldsymbol{\nabla}\theta_i^0, \label{apt16}\\
\nabla^2\boldsymbol{v}_0^1-Ra_c(\tau_1+a_1\tau_0-\theta_0^1)\boldsymbol{e}_z-\tau_0\boldsymbol{e}_z+\boldsymbol{\nabla}q &=\underbrace{-\frac{1}{\sigma }\sum_{i=1,2}\boldsymbol{v}_i^0\cdot\boldsymbol{\nabla}\boldsymbol{v}_i^0}_{balanced \; by\; pressure}. \label{apt19b}
\end{align}
which are coupled with the forced field equations
\begin{equation}\label{apt17}
-2\nabla^2\boldsymbol{v}_0^1
+2Ra_c\theta_0^1\boldsymbol{\nabla}\tau_0+\boldsymbol{\nabla}p=\nabla^2\boldsymbol{\phi}_1,
\end{equation}
\begin{equation}\label{apt19}
-2\nabla^2\theta_0^1+2\boldsymbol{v}_0^1\cdot\boldsymbol{\nabla}\tau_0=\phi_{13}.
\end{equation}
%
The forcing term in (\ref{apt16}) is
\begin{equation}\label{apt32}
\sum_{i=1}^2\boldsymbol{v}_i^0\cdot\boldsymbol{\nabla}\theta_i^0=\frac{\pi (A_1^2+A_2^2)}{6}\sin(2\pi z)
\end{equation}
and in (\ref{apt19b})
\begin{equation}\label{apt33}
\sum_{i=1}^2\boldsymbol{v}_i^0\cdot\boldsymbol{\nabla}\boldsymbol{v}_i^0=\frac{k\pi^2}{2}(A_1^2-A_2^2)\sin(2kx)\boldsymbol{e}_x+\frac{k^2\pi}{2}(A_1^2+A_2^2)\sin(2\pi z)\boldsymbol{e}_z
\end{equation}
(recall $k=\pi/\sqrt{2}$) so that  simply
\begin{equation}\label{apt34}
\boldsymbol{\phi}_1=\boldsymbol{v}_0^1={\bf 0}, \quad \theta_0^1=0, \quad q=c_1(A_1^2-A_2^2)\cos(2kx)+c_2 (A_1^2+A_2^2) \cos (2 \pi z)
\end{equation}
\begin{equation}
\& \quad \tau_1=c_2(A_1^2+A_2^2) \sin(2 \pi z)
\end{equation}
where $c_1, c_2$ and $c_3$ are specific constants. Finally, the leading order balance (which is at $O(\eps)$) in the balance parameter equation (\ref{apt6b}) is
\begin{equation}
-\frac{1}{a_1}\langle |\bnab \tau|^2 \rangle -\frac{1}{a_1} \langle \phi_{23}(1-z)\rangle-\frac{1}{Ra_c} \sum_{i=1,2} \langle |\boldsymbol{\nabla}\boldsymbol{v}^0_i|^2 \rangle=0
\label{balance}
\end{equation}
which relates $A_1^2+A_2^2$, $a_1$ and the higher order unknown $\boldsymbol{\phi}_2$.

\subsubsection{Next Order}

%
%
A further piece of information to identify the leading order fields comes from a solvability condition on the spectral constraint equations at next order ($(O(\eps^{3/2})$) which is
\begin{equation}\label{apt35}
-2\nabla^2\boldsymbol{v}_i^1
+2Ra_c\theta_i^1\boldsymbol{\nabla}\tau_0+\boldsymbol{\nabla}p=2a_1\nabla^2\boldsymbol{v}_i^0-2Ra_c\theta_i^0\boldsymbol{\nabla}\tau_1-2\theta_i^0\boldsymbol{\nabla}\tau_0,
\end{equation}
\begin{equation}\label{apt37}
-2\nabla^2\theta_i^1+2\boldsymbol{v}_i^1\cdot\boldsymbol{\nabla}\tau_0=-2\boldsymbol{v}_i^0\cdot\boldsymbol{\nabla}\tau_1.
\end{equation}
Formally, this has  two solvability conditions:
\begin{equation}\label{apt39}
\langle\boldsymbol{v}_i^0\cdot(2a_1\nabla^2\boldsymbol{v}_i^0-2Ra_c\theta_i^0\boldsymbol{\nabla}\tau_1-2\theta_i^0\boldsymbol{\nabla}\tau_0)-2\theta_i^0\boldsymbol{v}_i^0\cdot\boldsymbol{\nabla}\tau_1\rangle=0,\quad
i=1,2
\end{equation}
but they are equivalent since  $\tau_1$ is 1-dimensional (i.e. solely a function of $z$) with the resulting condition linking $a_1$ and $A_1^2+A_2^2$.  Only after $\boldsymbol{\phi}_2$ is found can $a_1$ and $A_1^2+A_2^2$ be fully determined from (\ref{balance}) and (\ref{apt39}).

The fields $(\boldsymbol{v}_i^1, \theta_i^1)$ are linearly dependent on $\boldsymbol{v}_i^0$, $\theta_i^0$ and so can be written as
\begin{equation}\label{apt39b}
(\boldsymbol{v}_1^1,\theta_1^1)=A_1(\, \mathcal{U}(z)\sin(kx)\boldsymbol{e}_x+\mathcal{W}(z)\cos(kx)\boldsymbol{e}_z,\mathcal {T}(z)\cos(kx)\,)
\end{equation}
\begin{equation}\label{apt39c}
(\boldsymbol{v}_1^2,\theta_1^2)=A_2(\, \mathcal{U}(z)\cos(kx)\boldsymbol{e}_x-\mathcal{W}(z)\sin(kx)\boldsymbol{e}_z,-\mathcal {T}(z)\sin(kx) \,)
\end{equation}
where $\mathcal{U}\neq U$, $\mathcal{W}\neq W$. These fields along with $(\boldsymbol{v}_i^0,\theta_i^0)$ drive the higher order equations governing further corrections  to the background fields and the forced fields. These are
\begin{equation}\label{apt40}
2\nabla^2\tau_2+\phi_{23}=\underbrace{2\sum_{i=1,2}\boldsymbol{v}_i^0\cdot\boldsymbol{\nabla}\theta_i^1+\boldsymbol{v}_i^1\cdot\boldsymbol{\nabla}\theta_i^0}_{driving \; term}.
\end{equation}
\begin{multline}\label{apt44}
\nabla^2\boldsymbol{v}_0^2
-Ra_c(a_0\tau_2+a_2\tau_0-\theta_0^2)\boldsymbol{e}_z+\boldsymbol{\nabla}q\\
=
\underbrace{([a_1 Ra_c+1] \tau_1 +a_1 \tau_0)\boldsymbol{e}_z
-\frac{a_1}{\sigma }\sum_{i=1}^2\boldsymbol{v}_i^0\cdot\boldsymbol{\nabla}\boldsymbol{v}_i^0}_{balanced \; by \;  pressure}-\underbrace{\frac{1}{\sigma }\sum_{i=1}^2\boldsymbol{v}_i^0\cdot\boldsymbol{\nabla}\boldsymbol{v}_i^1+\boldsymbol{v}_i^1\cdot\boldsymbol{\nabla}\boldsymbol{v}_i^0}_{driving \; term}=0.
\end{multline}
\begin{equation}\label{apt41}
-2\nabla^2\boldsymbol{v}_0^2
+2Ra_c\theta_0^2\boldsymbol{\nabla}\tau_0+\boldsymbol{\nabla}p_i=\nabla^2\boldsymbol{\phi}_2,
\end{equation}
\begin{equation}\label{apt43}
-2\nabla^2\theta_0^2+2\boldsymbol{v}_0^2\cdot\boldsymbol{\nabla}\tau_0=\phi_{23}.
\end{equation}
The apparent driving term $-a_1/\sigma\sum_{i=1}^2\boldsymbol{v}_i^0\cdot\boldsymbol{\nabla}\boldsymbol{v}_i^0$ can be balanced by the pressure term (see (\ref{apt33})\,) as can $([a_1 Ra_c+1]\tau_1+a_1\tau_0) \boldsymbol{e}_z$. Also importantly for what follows, $Ra_c a_2 \tau_0 \boldsymbol{e}_z$ in  (\ref{apt44}) can also be absorbed into the pressure term which means that $\tau_2$ and $\boldsymbol{\phi}_2$ do not depend on $a_2$ (this is crucial for the argument surrounding (\ref{solvability_1}) below).
This leaves the driving term for the 2D background temperature field
\begin{multline}\label{apt45b}
\sum_{i=1,2}\boldsymbol{v}_i^0\cdot\boldsymbol{\nabla}\theta_i^1+\boldsymbol{v}_i^1\cdot\boldsymbol{\nabla}\theta_i^0
=\frac{1}{2}(A_1^2+A_2^2)
\biggl(-kU\mathcal {T}-k\mathcal {U}\Theta+W\frac{d\mathcal {T}}{dz}+\mathcal {W}\frac{d\Theta}{dz} \biggr)
\\
+\frac{1}{2}(A_1^2-A_2^2)
\biggl(kU\mathcal {T}+k\mathcal {U}\Theta+W\frac{d\mathcal {T}}{dz}+\mathcal {W}\frac{d\Theta}{dz} \biggr)\cos(2kx).
\end{multline}
and  the driving term in Eq.(\ref{apt44}) for $\boldsymbol{v}_0^2$
\begin{multline}\label{apt45c}
\sum_{i=1}^2\boldsymbol{v}_i^0\cdot\boldsymbol{\nabla}\boldsymbol{v}_i^1+\boldsymbol{v}_i^1\cdot\boldsymbol{\nabla}\boldsymbol{v}_i^0
=\underbrace{\frac{1}{2}(A_1^2+A_2^2) \biggl(-kU\mathcal {W}-k\mathcal {U}W+\frac{d(W\mathcal {W})}{dz} \biggr) \boldsymbol{e}_z}_{balanced \; by \; pressure}\\+\frac{1}{2}(A_1^2-A_2^2)(2kU\mathcal {U}+W\frac{d\mathcal {U}}{dz}+\mathcal {W}\frac{dU}{dz})\sin(2kx)\boldsymbol{e}_x.
\end{multline}
Given the form of these driving terms, $\tau_2$ can be split into two parts: a 1D part which depends only on $z$ proportional to $A_1^2+A_2^2$, and a 2D part proportional to $A_1^2-A_2^2$ which has both $x$ and $z$ dependence whereas the remaining corrections $\boldsymbol{\phi}_2$, $\boldsymbol{v}_0^2$ and $\theta_0^2$ only have a 2D part proportional to $A_1^2-A_2^2$, so
\begin{align}
\tau_2                        &=\tau_2^{1D}(z)+\tau_2^{2D}(x,z):=(A_1^2+A_2^2)P_1(z)+(A_1^2-A_2^2)P_2(z)\cos(2kx), \label{apt46a}\\
\boldsymbol{\phi}_2 &=(A_1^2-A_2^2)\biggl[ G_1(z)\sin(2kx)\boldsymbol{e}_x+G_2(z)\cos(2kx)\boldsymbol{e}_z \biggr]. \label{apt46b}
\end{align}
(the expressions for $\boldsymbol{v}_0^2$ and $\theta_0^2$ are not needed in what follows and hence suppressed).  At this point $\boldsymbol{\phi}_2$ is now known as a function of $A_1^2-A_2^2$ and  even with the previously derived relations (\ref{balance}) and (\ref{apt39}), it is still not possible to identify $A_1$, $A_2$ and $a_1$ without further information about $A_1$ and $A_2$.  This comes
from solvability conditions at the next order of the spectral constraint ($O(\eps^{5/2})$).

Before pursuing this, we remark that the next order ($O(\eps^2)$) of the balance equation (\ref{apt6b}) involves the higher order  unknown $\boldsymbol{\phi}_3$ and so at this order $a_2$ is unspecified. In fact $a_2$ is also set by solvability conditions at $O(\eps^{5/2})$ of the spectral constraint to which we now turn.

\subsubsection{Solvability at $O(\eps^{5/2})$}

The spectral constraint equations at $O(\eps^{5/2})$ are
\begin{multline}\label{apt47}
-2\nabla^2\boldsymbol{v}_i^2
+2Ra_c\theta_i^2\boldsymbol{\nabla}\tau_0+\boldsymbol{\nabla}p_i=2a_1\nabla^2\boldsymbol{v}_i^1+2a_2\nabla^2\boldsymbol{v}_i^0-2Ra_c(\theta_i^0\boldsymbol{\nabla}\tau_2+\theta_i^1\boldsymbol{\nabla}\tau_1)\\
-2(\theta_i^0\boldsymbol{\nabla}\tau_1+\theta_i^1\boldsymbol{\nabla}\tau_0)-\frac{1}{\sigma}(\boldsymbol{\nabla}\boldsymbol{\phi}_2+\boldsymbol{\nabla}\boldsymbol{\phi}_2^T)\cdot\boldsymbol{v}_i^0
\end{multline}
\begin{equation}\label{apt49}
-2\nabla^2\theta_i^2+2\boldsymbol{v}_i^2\cdot\boldsymbol{\nabla}\tau_0=-2\boldsymbol{v}_i^0\cdot\boldsymbol{\nabla}\tau_2-2\boldsymbol{v}_i^1\cdot\boldsymbol{\nabla}\tau_1.
\end{equation}
Identifying the operator on the left hand side of (\ref{apt47})$-$(\ref{apt49}) as $\mathcal{L}$, then since it is self adjoint and annihilates the leading order fields $(\boldsymbol{v}_j^0,\theta_j^0)$ $j=1,2$,
there are  solvability conditions for $(\boldsymbol{v}_0^2,\theta_0^2)$
\begin{multline}\label{apt51}
\langle\boldsymbol{v}_j^0\cdot[2a_1\nabla^2\boldsymbol{v}_i^1+2a_2\nabla^2\boldsymbol{v}_i^0-2Ra_c(\theta_i^0\boldsymbol{\nabla}\tau_2+\theta_i^1\boldsymbol{\nabla}\tau_1)
-2(\theta_i^0\boldsymbol{\nabla}\tau_1+\theta_i^1\boldsymbol{\nabla}\tau_0)\\
-\frac{1}{\sigma}(\boldsymbol{\nabla}\boldsymbol{\phi}_2+\boldsymbol{\nabla}\boldsymbol{\phi}_2^T)\cdot\boldsymbol{v}_i^0]
-2\theta_j^0(\boldsymbol{v}_i^0\cdot\boldsymbol{\nabla}\tau_2+\boldsymbol{v}_i^1\cdot\boldsymbol{\nabla}\tau_1)\rangle=0.
\end{multline}
Taking $i=j$ (the $i \neq j$ conditions vanish trivially), this can be rearranged to
\begin{equation}
(A_1^2+A_2^2) {\rm Term_1}(i) + (A_1^2-A_2^2){\rm Term_2}(i)=0 \qquad i=1,2
\label{solvability}
\end{equation}
where
\begin{equation}
(A_1^2-A_2^2){\rm Term_2}(i) :=-\frac{1}{A_i^2}\biggl\langle
\frac{2}{\sigma} \boldsymbol{v}_i^0 \cdot \boldsymbol{\nabla}\boldsymbol{\phi}_2 \cdot \boldsymbol{v}_i^0
+2(Ra_c+1)\theta_i^0\boldsymbol{v}_i^0\cdot\boldsymbol{\nabla}\tau_2^{2D}
\biggr\rangle
\end{equation}
%
%
Crucially Term$_1(1)$=Term$_1(2)$ whereas Term$_2(1)$=$-$Term$_2(2)$ so that (\ref{solvability}) implies that
\begin{equation}
(A_1^2+A_2^2) {\rm Term_1}(1) =(A_1^2-A_2^2){\rm Term_2}(1)=0.
\label{solvability_1}
\end{equation}
The  unspecified coefficient $a_2$ only enters Term$_1$ and so is set by the condition this vanishes.
In contrast, there are no free constants in Term$_2$ which is non-zero in our computations (although we have been unable to prove this is always the case). In this situation, $A_1^2-A_2^2=0$ is forced instead which eliminates at a stroke all 2-dimensional fields in the bifurcation analysis. Consequently, the background flow field remains zero and the background temperature field stays 1D after the first bifurcation.

\subsection{Subsequent bifurcations}
Now we consider subsequent bifurcations to establish that if $\tau=\tau(z),\boldsymbol{\phi}=0$ exists before then that situation persists after the bifurcation. The approach is inductive: assume $\tau=\tau(z),\boldsymbol{\phi}=0$ after $m$ bifurcations and consider the $(m+1)^{th}$ bifurcation at $Ra=Ra_c^{(m+1)}$ where two new neutral modes appear so that there are now $2(m+1)$ critical modes in the spectral constraint. Defining $\eps:=Ra-Ra_c^{(m+1)}$ we expand:
\begin{align}\label{next1}
(\tau, \boldsymbol{\phi})       & =(\;\tau_0(z)+\eps \tau_1(x,z)+\dots,\; \eps \boldsymbol{\phi}_0(x,z)+\dots) \\
(\boldsymbol{v}_0, \theta_0) & =(\;\eps \boldsymbol{v}_0^0+\dots,\; \eps \theta_0^0+\dots)\\
a &=a_0+\eps a_1+\dots
\end{align}
\begin{equation}
(\boldsymbol{v}_i, \theta_i)=\biggl\{ \begin{array}{lll}
(\boldsymbol{v}_i^0+\eps \boldsymbol{v}_i^1+\dots,
& \theta_i^0+\eps \theta_i^1+\dots)  &  \quad i=1,2,\dots,2m \\
(\eps^{1/2}\boldsymbol{v}_i^0+\eps^{3/2}\boldsymbol{v}_i^1+\dots,
& \eps^{1/2}\theta_i^0+\eps^{3/2}\theta_i^1+\dots) &  \;\; \; i=2m+1,2m+2.
\end{array} \biggr.
\end{equation}
where the  leading fields $\tau_0(z)$, $(\boldsymbol{v}_i^0,\theta_i^0)$ ($i=1,\ldots 2m+2$) and $a_0$ are all known. In particular, the $i^{th}$ wavenumber $k_i$ ($i=1,2,\dots,m$), is associated with two modes which, to leading order, are
\begin{equation}\label{next8a}
(\boldsymbol{v}_{2i-1}^0, \theta_{2i-1}^0)=A_i(\,U_i(z)\sin(k_ix)\boldsymbol{e}_x+W_i(z)\cos(k_ix)\boldsymbol{e}_z,\Theta_i(z)\cos(k_ix) \,)
\end{equation}
and
\begin{equation}\label{next8b}
(\boldsymbol{v}_{2i}^0,\theta_{2i}^0)=B_i(\, U_i(z)\cos(k_ix)\boldsymbol{e}_x-W_i(z)\sin(k_ix)\boldsymbol{e}_z,-\Theta_i(z)\sin(k_ix)\,)
\end{equation}
where $A_i^2=B_i^2$ for $i=1,2,...,m$. The two new modes emerging at the $(m+1)^{th}$ bifurcation point can be assumed to have  the following general 3D form:
\begin{align}\label{next9a}
\boldsymbol{v}^0_{2m+1}&=A_{m+1}\left(\begin{array}{c}
                                  U_{m+1}(z)\sin(\boldsymbol{k}_{m+1}\cdot\boldsymbol{x}) \\
                                  V_{m+1}(z)\sin(\boldsymbol{k}_{m+1}\cdot\boldsymbol{x}) \\
                                  W_{m+1}(z)\cos(\boldsymbol{k}_{m+1}\cdot\boldsymbol{x})
                                \end{array}\right)\\
\theta_{2m+1}^0&=A_{m+1}\Theta_{m+1}(z)\cos(\boldsymbol{k}_{m+1}\cdot\boldsymbol{x}),
\end{align}
and
\begin{align}\label{next9a}
\boldsymbol{v}^0_{2m+2}&=B_{m+1}\left(\begin{array}{c}
                                  U_{m+1}(z)\cos(\boldsymbol{k}_{m+1}\cdot\boldsymbol{x}) \\
                                  V_{m+1}(z)\cos(\boldsymbol{k}_{m+1}\cdot\boldsymbol{x}) \\
                                  -W_{m+1}(z)\sin(\boldsymbol{k}_{m+1}\cdot\boldsymbol{x})
                                \end{array}\right)\\
\theta_{2m+2}^0&=-B_{m+1}\Theta_{m+1}(z)\sin(\boldsymbol{k}_{m+1}\cdot\boldsymbol{x}),
\end{align}
where $\boldsymbol{k}_{m+1}=(k_x,k_y,0)$. The objective in what follows is to show that  $A_{m+1}^2=B_{m+1}^2$ after the bifurcation so that the optimization problem remains 1D.

At leading order ($O(\epsilon)$) in the forced field and background field equations
\begin{equation}\label{next10}
-2a_0\nabla^2\boldsymbol{v}_0^0+2Ra_c^{m+1}\theta_0^0\boldsymbol{\nabla}\tau_0+\boldsymbol{\nabla}p-a_0\nabla^2\boldsymbol{\phi}_0={\bf 0},
\end{equation}
\begin{equation}\label{next12}
-2\nabla^2\theta_0^0+2\boldsymbol{v}_0^0\cdot\boldsymbol{\nabla}\tau_0+2\boldsymbol{\phi}_0\cdot\boldsymbol{\nabla}\tau_0+a_0\phi_{03}=0.
\end{equation}
\begin{multline}\label{next13}
a_0\nabla^2\boldsymbol{v}_0^0-Ra_c^{m+1}(a_0\tau_1+a_1\tau_0+a_0\theta_0^0)\boldsymbol{e}_z-a_0\tau_0\boldsymbol{e}_z-2Ra_c^{m+1}\theta_0^0\boldsymbol{\nabla}\tau_0+\boldsymbol{\nabla}q\\
+\frac{a_0}{\sigma }\sum_{i=1}^{2m} \biggl(\boldsymbol{v}_i^0\cdot\boldsymbol{\nabla}\boldsymbol{v}_i^1+\boldsymbol{v}_i^1\cdot\boldsymbol{\nabla}\boldsymbol{v}_i^0\biggr)
=-\underbrace{\frac{a_1}{\sigma }\sum_{i=1}^{2m}\boldsymbol{v}_i^0\cdot\boldsymbol{\nabla}\boldsymbol{v}_i^0}_{1D}
-\underbrace{\frac{a_0}{\sigma }\sum_{i=2m+1}^{2m+2}\boldsymbol{v}_i^0\cdot\boldsymbol{\nabla}\boldsymbol{v}_i^0}_{driving \;term},
\end{multline}
\begin{multline}\label{next15}
-2\nabla^2\tau_1-a_0\phi_{03}+2\sum_{i=1}^{2m} \biggl( \boldsymbol{v}_i^0\cdot\boldsymbol{\nabla}\theta_i^1+\boldsymbol{v}_i^1\cdot\boldsymbol{\nabla}\theta_i^0 \biggr)=
-\underbrace{2\sum_{i=2m+1}^{2m+2}\boldsymbol{v}_i^0\cdot\boldsymbol{\nabla}\theta_i^0}_{driving \; term}.
\end{multline}
where again it is implicit that $\boldsymbol{v}_0$, $\boldsymbol{v}_i$ and $\boldsymbol{\phi}$ are incompressible fields.
The spectral constraint for modes $i=1,2,...,2m$ at $O(\eps)$ and for modes $i=2m+1,2m+2$ at $O(\eps^{3/2})$ is
\begin{multline}\label{next16}
-2a_0\nabla^2\boldsymbol{v}_i^1+2Ra_c^{m+1}\theta_i^1\boldsymbol{\nabla}\tau_0+\boldsymbol{\nabla}p=
2a_1\nabla^2\boldsymbol{v}_i^0-\frac{a_0}{\sigma}\boldsymbol{v}_i^0\cdot(\boldsymbol{\nabla}\boldsymbol{\phi}_0+\boldsymbol{\nabla}\boldsymbol{\phi}_0^{\texttt{T}})\\
-2Ra_c^{m+1}\theta_i^0\boldsymbol{\nabla}\tau_1-2\theta_i^0\boldsymbol{\nabla}{\tau}_0,
\end{multline}
\begin{equation}\label{next18}
-2\nabla^2\theta_i^1+2\boldsymbol{v}_i^1\cdot\boldsymbol{\nabla}\tau_0=-2\boldsymbol{v}_i^0\cdot\boldsymbol{\nabla}\tau_1.
\end{equation}

The  system of equations (\ref{next10})-(\ref{next18}) is linear in $\boldsymbol{v}_0^0$, $\theta_0^0$, $\boldsymbol{\phi}_0$, $\boldsymbol{v}_i^1$  and
$\theta_i^1$ ($i=1,2,...,2m+2$). The emergent critical modes at $Ra_c^{m+1}$ give rise to the new  driving terms in (\ref{next13}) \& (\ref{next15})
\begin{align}
\sum_{i=2m+1}^{2m+2} \boldsymbol{v}_i^0 \cdot \boldsymbol{\nabla}\boldsymbol{v}_i^0
&=\tfrac{1}{2} (A_{m+1}^2+B_{m+1}^2) \boldsymbol{\nabla}[W^2_{m+1}(z)] \nonumber \\
&   +\tfrac{1}{2}(A_{m+1}^2-B_{m+1}^2) \left[ \begin{array}{c}
-U_{m+1}\frac{dW_{m+1}}{dz}+\frac{dU_{m+1}}{dz} W_{m+1} \\
-V_{m+1}\frac{dW_{m+1}}{dz}+\frac{dV_{m+1}}{dz} W_{m+1} \\
0
\end{array} \right]
\sin (2 \boldsymbol{k}_{m+1} \cdot \boldsymbol{x} )\\
\sum_{i=2m+1}^{2m+2}\boldsymbol{v}_i^0 \cdot \boldsymbol{\nabla}\theta_i^0 &=
\tfrac{1}{2} (A_{m+1}^2+B_{m+1}^2) \frac{d (W_{m+1} \Theta_{m+1}) }{dz} \nonumber\\
&+\tfrac{1}{2}(A_{m+1}^2-B_{m+1}^2) \biggl( W_{m+1} \frac{d \Theta_{m+1}}{dz}-\frac{dW_{m+1}}{dz} \Theta_{m+1} \biggr) \cos (2 \boldsymbol{k}_{m+1} \cdot \boldsymbol{x} )
\end{align}
which have a 1D part proportional to $A_{m+1}^2+B_{m+1}^2$ and a non-1D part proportional to $A_{m+1}^2-B_{m+1}^2$. {\em If} $A^2_{m+1}=B^2_{m+1}$ then $\tau_1=\tau_1(z)$ and
$\boldsymbol{\phi}={\bf 0}$ using the arguments of  section \ref{fbp}. Hence for $A^2_{m+1} \neq B^2_{m+1}$ and using equation (\ref{next15}), we can assume a solution structure of the form
\begin{align}\label{next21}
\tau_1 &=\underbrace{\sum_{i=1}^{m+1}(A^2_{i}+B^2_{i})P_i(z)}_{\tau_1^{1D}}+\underbrace{(A^2_{m+1}-B^2_{m+1}) \biggl(Q(z) \cos(2\boldsymbol{k}_{m+1}\cdot\boldsymbol{x})}_{\tau_1^{2D}}+\tau_1^*(x,z) \biggr),\\
\boldsymbol{\phi}_0 & = (A^2_{m+1}-B^2_{m+1})\left(
\left[
\begin{array}{c}
G_1(z)  \sin(2\boldsymbol{k}_{m+1}\cdot\boldsymbol{x}) \\
G_2(z) \sin(2\boldsymbol{k}_{m+1}\cdot\boldsymbol{x})\\
G_3(z) \cos(2\boldsymbol{k}_{m+1}\cdot\boldsymbol{x})
\end{array}
\right]+\boldsymbol{\Phi}^*(x,z) \;\right)
\end{align}
where $\tau_1^*(x,z)$ and $\boldsymbol{\Phi}^*(x,z)$  collect all the other wavenumber dependence on $x$ in $\tau_1$ and $\boldsymbol{\phi}_0$ respectively (this is unimportant in what follows).
The key is then examining the solvability conditions
\begin{multline}\label{next22}
\biggl\langle\boldsymbol{v}_j^0\cdot\left(2a_1\nabla^2\boldsymbol{v}_i^0-\frac{a_0}{\sigma}\boldsymbol{v}_i^0\cdot(\boldsymbol{\nabla}\boldsymbol{\phi}_0+\boldsymbol{\nabla}\boldsymbol{\phi}_0^\texttt{T})
-2Ra_c^{m+1}\theta_i^0\boldsymbol{\nabla}\tau_1-2\theta_i^0\boldsymbol{\nabla}{\tau}_0\right) \biggr.\\
\biggl.-2\theta_i^0\boldsymbol{v}_j^0\cdot\boldsymbol{\nabla}\tau_1\biggr\rangle=0
\end{multline}
($i,j,=1,\ldots m+1$) on the corrections $\boldsymbol{v}_i^1$ to all the critical modes $\boldsymbol{v}_i^0$ of the spectral constraint. These set the amplitudes $A_i^2\,(=B_m^2)$ ($i=1,\ldots m$) and $(A_{m+1},B_{m+1})$ ($a_1$ is determined by the balance parameter equation (\ref{apt6b}) at $O(\eps)$).

To establish that the background fields stay 1D, it is sufficient to focus on the solvability conditions for the new critical modes ($i=2m+1$ and $2m+2$). Here, the solvability condition is explicitly
\begin{multline}\label{next24}
-a_1\int^1_0 \boldsymbol{k}_{m+1}^2(U_{m+1}^2+V_{m+1}^2+W_{m+1}^2)+\left(\frac{dU_{m+1}}{dz}\right)^2+\biggl(\frac{dV_{m+1}}{dz}\biggr)^2+\left(\frac{dW_{m+1}}{dz}\right)^2dz\\-\int^1_0W_{m+1}\Theta_{m+1}\frac{d\tau_0}{dz}dz
+\sum_{j=1}^{m+1} \biggl[(A_{j}^2+B_{j}^2)(Ra_c^{m+1}+1)\int^1_0W_j\Theta_j\frac{dP_j}{dz}dz \biggr]\\+ (A_{m+1}^2-B_{m+1}^2){\rm Term}(i)=0
\end{multline}
where
\begin{multline}\label{next24}
(A_{m+1}^2-B_{m+1}^2){\rm Term}(i):=-\frac{2}{c}\biggl\langle\frac{a_0}{\sigma}\boldsymbol{v}_i^0\boldsymbol{\nabla}\boldsymbol{\phi}_0\cdot\boldsymbol{v}_i^0+(Ra_c^{m+1}+1)\theta_i^0\boldsymbol{v}_i^0\cdot\boldsymbol{\nabla}\tau_1^{2D} \biggr\rangle,
\end{multline}
with $c:=A_{m+1}^2$ for $i=2m+1$ or  $B_{m+1}^2$ for $i=2m+2$. Crucially, it is straightforward to show that
\begin{equation}
{\rm Term}(2m+1)=-{\rm Term}(2m+2)
\end{equation}
so that, as in subsection (\ref{fbp}), we must have
\begin{equation}
(A_{m+1}^2-B_{m+1}^2){\rm Term}(i)=0
\label{term(2m+1)}
\end{equation}
Our computations indicate ${\rm Term}(2m+1) \neq 0$ (although, as in subsection (\ref{fbp}), we have been unable to prove this in general) forcing $A_{m+1}^2=B_{m+1}^2$. This forces $\tau_1=\tau_1(z)$ and $\boldsymbol{\phi}_0={\bf 0}$ so that the optimal solution stays 1D after the $(m+1)^{th}$ ($m \geq 0$) bifurcation if it is of this form before. Taken together with the first bifurcation analysis in subsection (\ref{fbp}), this means that the optimal solution is 1D for all $Ra$ and so, surprisingly, there is {\em no} benefit of imposing the full momentum and heat balances in the upper bound problem.

%
%
\section{Adding a background velocity field to \cite{Wen}}
%

Following the success of adding an enstrophy constraint in 2D stress-free convection \cite{Wen}, an interesting question is whether adding a 1-D background velocity field by using the decomposition
\begin{equation}\label{vec1}
 \boldsymbol{u}(x,z)=\phi(z)\boldsymbol{e}_x+\boldsymbol{v}(x,z),\quad
 T(x,z)=\tau(z)+\theta(x,z)
\end{equation}
would improve the bound further since this imposes additional information from the Navier-Stokes equations. To maximize the heat flux, the Lagrangian
\begin{equation}\label{vec4}
 Nu=\langle|\boldsymbol{\nabla}T|^2\rangle-\frac{a}{\sigma Ra}\langle\boldsymbol{v}\cdot \boldsymbol{\mathcal{N}}\rangle-\frac{b}{\sigma Ra}\langle \boldsymbol{\omega} \cdot \boldsymbol{\nabla} \times \boldsymbol{\mathcal{N}})\rangle-2\langle\theta \,\mathcal {H}\rangle
\end{equation}
is constructed where $\boldsymbol{\omega}=\omega(x,z) \boldsymbol{e}_y:=(\boldsymbol{\nabla}\times\boldsymbol{v})$.
After some  integration by parts, judicious use of boundary conditions, and building in the fact that $Nu=1+\langle wT\rangle$, this leads to the expression
\begin{align}
(1-a) Nu +a =& \langle|\tau'|^2\rangle
-\biggl\langle
|\nabla \theta|^2+2 \theta v_3 \tau'+\frac{a}{\sigma Ra}v_1 v_3 \phi' +\frac{b}{\sigma Ra} v_3 \omega \phi''-b \omega \frac{\partial \theta}{\partial x}
\nonumber \\
& \qquad +\frac{a}{Ra}\biggl(|\nabla \boldsymbol{v}|^2-v_1 \phi'' \biggr)
+\frac{b}{Ra}\biggl( |\nabla \omega|^2-\omega \phi''' \biggr)
\biggr\rangle.
\end{align}
The two linear terms in the second line of this expression - $ v_1 \phi''$ and $\omega \phi''' $ - mean that optimization over the fluctuation fields  $\boldsymbol{v}$ and $\theta$ will give rise to a non-zero contribution to be added to $ \langle|\tau'|^2\rangle$. This complication can be avoided (or rather made more explicit) by defining a shifted variable
\begin{equation}
\hat{\boldsymbol{v}}:=\boldsymbol{v}+\half \phi(z) \boldsymbol{e}_x
\end{equation}
which is possible if  $\boldsymbol{v}$ and $\phi$ are both assumed to satisfy (natural) homogeneous boundary conditions and allows both linear terms to be absorbed into perfect squares. As a result of this, the expression becomes
\begin{align}\label{vec7}
 Nu=\frac{1}{1-a}(\langle|\tau'|^2\rangle-a)+&\frac{a}{4(1-a)Ra}\langle|\phi'|^2\rangle +\frac{b}{4(1-a)Ra}\langle|\phi''|^2\rangle
\nonumber \\
&\qquad - \frac{1}{1-a}\mathscr{G}(\hat{\boldsymbol{v}},\omega,\theta;\tau,\phi, a, b, Ra, \sigma)
\end{align}
where
\begin{equation}\label{vec8}
\mathscr{G}:=a\langle \frac{1}{Ra}|\nabla \hat{\boldsymbol{v}}|^2+\frac{1}{\sigma Ra}\hat{v}_1 \hat{v}_3\phi'\rangle
+
b\langle \frac{1}{\sigma Ra}\omega \hat{v}_3\phi''+{\frac{1}{Ra}}|\boldsymbol{\nabla}\hat{\omega}|^2
-\hat{\omega}\frac{\partial\theta}{\partial x}\rangle
+\langle|\boldsymbol{\nabla}\theta|^2+2\theta \hat{v}_3\tau'\rangle \nonumber
\end{equation}
is a purely quadratic functional of $\hat{\boldsymbol{v}}$,
$\hat{\omega}:=\boldsymbol{e}_y \cdot \boldsymbol{\nabla} \times \hat{\boldsymbol{v}}=\omega+\half \phi'$ and $\theta$. Provided the background fields $\tau$ and $\phi$ are chosen such that $\mathscr{G} \geq 0$ for all permissable $\hat{\boldsymbol{v}}$, $\hat{\omega}$ and $\theta$, then a bound follows on $Nu$.  Now it is clear that:  1) the objective functional is convex in the background fields  and 2) the set of allowable background fields is convex (if $(\tau_1,\phi_1)$ and $(\tau_2,\phi_2)$ ensure $\mathscr{G} \geq 0$ so does $(\tau,\phi)=\lambda (\tau_1, \phi_1)+(1-\lambda)(\tau_2,\phi_2)$ with $0\leq \lambda \leq 1$). This implies that the optimizer is unique and is attained for $(\tau,\phi)=(\tau,0)$ i.e. the background velocity field vanishes indicating that the extra information this folds into the optimization is, in fact, unimportant. Physically, a bifurcation analysis shows that the fluctuation fields are always such as to produce zero Reynolds stress so that no background field is generated.

\begin{figure}
 \centering
  \scalebox{0.85}[0.85]{\includegraphics[bb=0 0 218 201]{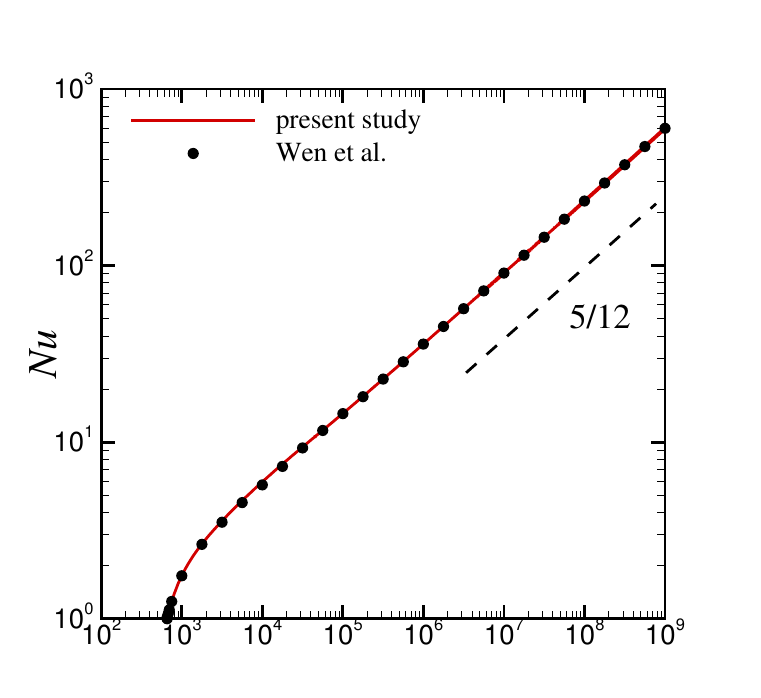}}
   \scalebox{0.85}[0.85]{\includegraphics[bb=0 0 218 201]{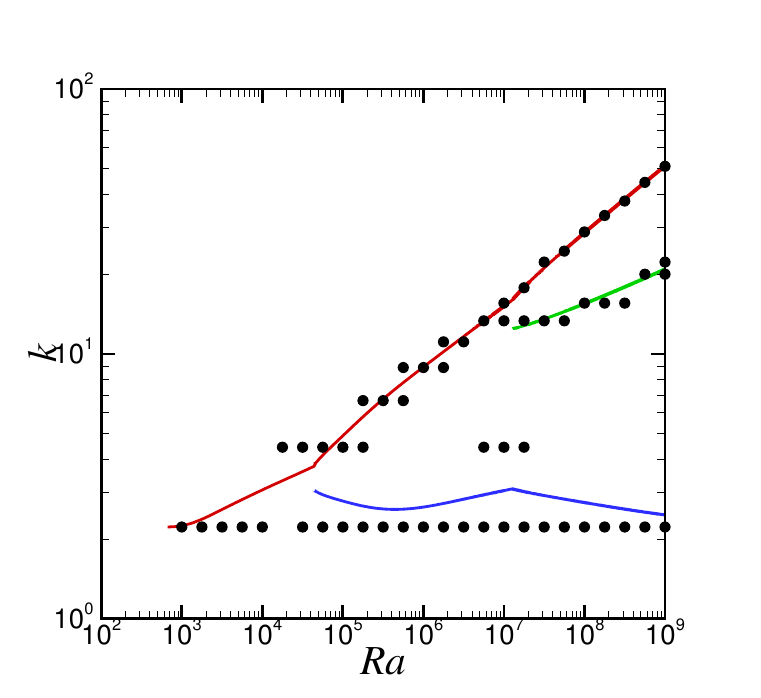}}
 \caption{\label{fig12} Left panel: the upper bound of $Nu$ vs. the Rayleigh number $Ra$, and $Nu\le 0.107a^{5/12}$ in the asymptotic ultimate regime. right panel: the bifurcation diagram of critical wavenumbers $k_m$ vs. the Rayleigh number (solid found here for $L=\infty$; dots from \cite{Wen} for fixed $L=2 \sqrt{2}$). The prefactor predicted $0.107$ is very slightly higher than \cite{Wen}'s $0.106$ (data courtesy of Baole Wen).}
\end{figure}

A numerical solution shown in figure \ref{fig12} using Newton's method on the Euler-Lagrange equations in an infinitely long domain confirms that $\phi=0$ as does a bifurcation analysis developed in the previous section. The bound compares well with the earlier results of \cite{Wen} who considered a fixed domain of $L=2 \sqrt{2}$ indicating further that the bound is not that sensitive to the domain size. The fashion in which the necessarily discretized critical modes found by  \cite{Wen} cluster around the (continuous) optimal wave numbers in our study confirms this conclusion.

%
%

%
%
\section{Discussion}
%

This paper has revisited the optimal heat transport problem  in two-dimensional
Rayleigh-B\'enard convection with stress-free boundary conditions using an extended background method. The key novelty has been to consider background temperature and velocity fields whose dimensional dependence matches that of the physical problem (so 2D here). This situation needs a  reformulation in the way the variational equations are solved which has the significant consequence of breaking any link between the optimal fields which emerge and a single physical temperature and velocity field. In particular, this means that the optimal fields do not obviously satisfy the heat equation even though that is explicitly imposed  when the background temperature field is allowed to be fully 2D. This is due to the spectral constraint (that ensures a bound) which means the optimal bound found does not correspond with the highest  stationary point of the Lagrangian  (i.e. the Euler-Lagrange equations are not all satisfied) but is strictly above it.
In other words, there is a gap between the highest heat flux attained by a steady solution of the governing equations imposed  and the best (lowest) bound because of the additional spectral constraint. Unfortunately and importantly,  this means that there is no direct connection between the optimal solution in the background method built around the steady governing equations and a steady solution of the governing equations (here the Boussinesq equations but clearly more generally true).  This realisation removes the possibility, for example, that the simple 2D roll solution computed by Waleffe et al. (2015) could actually be the optimal solution to the background bounding problem. It now seems clear that it would be spectrally unstable.

In revisiting  the exact 2D Rayleigh-Benard problem treated by Hassanzadeh et al. (2014), we have shown that their maximal heat flux result is only a global maximum up to $Ra\le Ra_c:=4468.8$. Beyond this, the optimal solution complexifies over a given spatial domain. If this domain is extended, the optimal becomes increasingly 1D. Removing the symmetry imposed by Hassanzadeh et al. (2014) and reinstating translational invariance in the horizontal direction by making the domain unbounded,  the optimal solution is then provably just 1D and the classic scaling result of $Nu \sim Ra^{1/2}$ is recovered albeit with the larger numerical coefficient
of 0.055 as opposed to the already known 0.026 (PK03) for non-slip boundary conditions. The conclusion is then that imposing the full heat equation in the bounding calculation does {\em not} improve (lower) the bound over that obtained using the horizontally-averaged heat equation. We then considered adding extra information from the momentum equation to  the upper bounding problem by introducing a background velocity field $\bphi(x,z)$. Now the optimization problem is no longer convex and we use an inductive bifurcation analysis to show that if $\bphi={\bf 0}$ before a bifurcation then it remains ${\bf 0}$ after it too. This means that the continuous branch of optimals found by branch tracking out of the energy stability point always has $\bphi={\bf 0}$. Noting the  caveats that a) it's not  impossible that there is an unconnected branch of optimals with $\bphi \neq {\bf 0}$ and b) $Term(2m+1)$ in (\ref{term(2m+1)}) could serendipitiously vanish at a subsequent bifurcation beyond our calculations, this strongly suggests the surprising result that imposing the full Boussinesq equations does {\em not} improve the bound over that obtained using the horizontally-averaged Boussinesq equations.

The `take-home' message from this study is that the background method of seeking an upper bound on heat flux in Rayleigh-Benard convection has been exhausted with disappointingly  {\em no} improvement possible over the minimal choice of a 1D background temperature field originally made in 1996 by Doering and Constantin. It's hard not to imagine this realisation  also generalising to the analogous background formulations for shear flows too, e.g. plane Couette flow (Doering \& Constantin, 1992), channel flow (Constantin and Doering, 1995) and pipe flow (Plasting \& Kerswell, 2005).  Simply extending the definition of the background fields ostensibly folds in more information from the governing equations but not in a fruitful way.  However, it seems generating extra information by {\em differentiating} the governing equations  can help. Whitehead \& Doering (2011) (see also Wen et al. 2015) used an extra vorticity constraint  to significantly lower the bound from $Nu \sim Ra^{1/2}$ to $Nu \sim Ra^{5/12}$ but only in the 2D situation with stress-free boundary conditions. Interestingly, this approach can be inverse-engineered into the form of  background method by loosening the connection between the Lagrange multiplier $\bnu(\bx,t)$ and the velocity field $\bu(\bx,t)$ from $\bu(\bx,t)-\bnu(\bx,t)=\phi(z) \hat{\bx}$ to
$$
\bu(\bx,t)-\bnu(\bx,t)=\phi(z) \hat{\bx}+c\boldsymbol{\nabla} \times \boldsymbol{\nabla} \times \bu(\bx,t)
$$
where $c$ is a new scalar Lagrange multiplier imposing the global vorticity constraint
$$
\langle \boldsymbol{\nabla} \times \bu \cdot \boldsymbol{\nabla} \times (\N)_s \rangle=0.
$$
This clearly extracts something more from  the governing equations than just taking projections. Maybe there is some mileage in exploring this but a shortage of boundary conditions is the usual impediment to this approach. Looking ahead, an emergent Sum-of-Squares approach to bounding (e.g. Fantuzzi et al. 2016, Goluskin \& Fantuzzi 2019) offers far greater potential for progress since it extends the quadratic constraints used here to more general polynomials albeit at the expense of a fully numerical approach.

\vspace{0.75cm}
\noindent
Acknowledgements: The authors are very grateful to Andre Souza and Charlie Doering for helpful discussions and sharing their recent preprint (Souza et al. 2019). The authors also acknowledge the support of EPSRC under grant EP/P001130/1.

%
%
\appendix
\section{Time stepping for a 2D background temperature field  in 2D}

Here we show that the time-marching method of Wen et al. (2015) is not guaranteed to have the optimal solution as the unique steady attractor when  $\tau=\tau(x,z)$ has the same {\em spatial} dimensionality as the physical temperature field $T(x,z,t)$. Time-stepping would work, however, for a 2-dimensional $\tau(x,z)$ in a 3-dimensional problem. To explain this, we  revisit the proof of Wen et al. (2015). The time-stepping approach consists of adding time derivatives for $\theta$, $\nabla^2 \psi$ and $\tau$ to the left hand sides of (\ref{EL_1})-(\ref{EL_3}) respectively.
%
%
Small disturbances $(\theta', \boldsymbol{u}', \tau', p')$ on top of a solution to the Euler-Lagrange equations, $(\theta,\boldsymbol{u}, \tau, p)$, then evolve according to the following equations
\beqa
\frac{\partial\theta'}{\partial t}              &&=  \nabla^2\theta'-J(\tau',\psi)-J(\tau,\psi')  ,         \label{T_1}\\
\frac{\partial \nabla^2 \psi'}{\partial t} &&= \frac{a}{Ra}\nabla^4\psi'-J(\tau',\theta)-J(\tau,\theta') ,     \label{T_2}\\
\frac{\partial\tau'}{\partial t}                 &&=  \nabla^2\tau' - J(\theta', \psi)- J(\theta,\psi')     \label{T_3}
\eeqa
at fixed balance parameter $a$. Then $\langle \, \theta' (\ref{T_1})-\psi' (\ref{T_2})+\tau' (\ref{T_3}) \,\rangle$ gives
\begin{equation}\label{stability}
\frac{\partial}{\partial t}
 \frac{1}{2}\langle{\theta'}^2+|\boldsymbol{\nabla} \psi'|^2+{\tau'}^2\rangle=-\langle|\boldsymbol{\nabla}\tau'|^2\rangle
 -\underbrace{
\langle \frac{a}{Ra}|\nabla^2 \psi'|^2+|\boldsymbol{\nabla}\theta'|^2+2\theta' J(\tau, \psi') \rangle
}_{\mathscr{G}}.
\end{equation}
In the 1-dimensional background field case, $\tau=\tau(z)$, (\ref{T_3}) becomes
\begin{equation}
\frac{\partial\tau'}{\partial t}   -\frac{\partial^2 \tau'}{\partial z^2}= - \overline{J(\theta', \psi)}- \overline{J(\theta,\psi')}
\label{1D}
\end{equation}
(where the overbar represents averaging over $x$) and the possible fluctuation fields can, after a Fourier transform, be assumed to have a specific wavenumber in $x$.  There are then two types of fluctuation fields: 1) those with wavenumbers which don't overlap with those in the optimal solution ($\lambda <0$ in the spectral constraint) and therefore do not generate any concomitant disturbance $\tau'$, and 2) those which do have a non-vanishing $\tau'$  but necessarily have $\lambda=0$ (the optimal solution is unique  for any {\em given} balance parameter $a \in (0,1)$ by the same arguments presented in the main text and, by construction, includes any fluctuation fields $(\theta, \psi)$ which are neutral in the spectral constraint). In both cases, the fluctuation fields have to decay, in the former case because $\lambda <0$ and in the latter through the $\tau'$ component generated in (\ref{1D}). The unique solution is therefore an attractor but the key step is proving that it is the only such. This follows by realising that if a solution to the Euler-Lagrange equations does not satisfy the spectral constraint, then there is an unstable eigenfunction of the linear time-stepping operator defined in (\ref{T_1})-(\ref{T_3}) which consists of the fluctuation field $(\theta',\psi')$  which makes $\mathscr{G} <0$. This is because a fluctuation field with $\lambda \neq 0$ does not overlap under $x$-averaging with the underlying state and so does not generate a $\tau'$ component via (\ref{1D}). This argument can clearly be extended to 2-dimensional $\tau(x,z)$ in 3-dimensional Rayleigh-Benard convection
since orthogonality in $x$ is replaced by orthogonality of $y$ but breaks down for 2-dimensional Rayleigh-Benard convection. In the latter situation, fluctuation fields which violate the spectral constraint will generate a $\tau'$ component via (\ref{T_3}) and may not then represent a growing eigenfunction for the time stepping procedure. The implication of this is that some saddles of $\opL$ may also be local attractors so if the time-stepping procedure leads to a steady state it is not guaranteed to be the optimal solution. Preliminary numerical tests  demonstrated this multistability with the final steady state depending on the initial condition used.

\bibliographystyle{jfm}

\begin{thebibliography}{}
\bibitem[Ahlers \etal (2009)]{Ahlers} \textsc{Ahlers G.}, \textsc{Grossmann S.} and \textsc{Lohse D.} 2009 Heat transfer and large scale dynamics in turbulent Rayleigh-Benard convection. \emph{Rev. Mod. Phys.} \textbf{81}, 503.

\bibitem[Busse (1969)]{Busse} \textsc{Busse F.H.} 1969 On Howard's upper bound for heat transport by turbulent convection. \emph{J. Fluid Mech.} \textbf{37}, 457.

\bibitem[Busse (1978)]{Busse78} \textsc{Busse F.H.} 1978 The optimum theory of turbulence \emph{Adv. Appl. Mech.} \textbf{18}, 77.

\bibitem[Constantin \& Doering  (1995)]{Doering3} \textsc{Constantin P.} and \textsc{Doering, C.R.} 1995 Variational bounds on energy
dissipation in incompressible flows: II. channel flow. \emph{Phys. Rev. E} \textbf{51}, 3192-3198.

\bibitem[Doering \& Constantin (1992)]{Doering1} \textsc{Doering C.R.} and  \textsc{Constantin P.} 1992 Energy dissipation in shear
driven turbulence. \emph{Phys. Rev. Lett.} \textbf{69}, 1648.

\bibitem[Doering \& Constantin (1994)]{Doering2} \textsc{Doering C.R.} and \textsc{Constantin P.} 1994 Variational bounds on energy
dissipation in incompressible flows: shear flow. \emph{Phys. Rev. E} \textbf{49}, 4087.

\bibitem[Doering \& Constantin (1996)]{Doering4} \textsc{Doering C.R.} and \textsc{Constantin P.} 1996 Variational bounds on energy
dissipation in incompressible flows. III. convection. \emph{Phys. Rev. E} \textbf{53}, 5957.

\bibitem[Fantuzzi (2018)]{Fantuzzi} \textsc{Fantuzzi, G.} 2018 Construction of optimal background fields using semidefinite programming \emph{PhD thesis} \emph{Imperial College}.

\bibitem[Fantuzzi et al. (2016)]{Fantuzzi_etal} \textsc{Fantuzzi, G.} , \textsc{Goluskin, D.}, \textsc{ Huang, D.} and \textsc{Chernyshenko S.I.} 2016 Bounds for deterministic and stochastic dynamical systems using sum-of-squares optimization \emph{SIAM J. App. Dyn. Sys.} \textbf{15}, 1962-1988.

\bibitem[Goluskin \& Fantuzzi (2019)]{Goluskin} \textsc{Goluskin, D.} and  \textsc{Fantuzzi, G.} 2019 Bounds on mean energy in the Kuramoto-Sivashinsky equation computed using semidefinite programming  \emph{Nonlinearity} \textbf{32}, 1705-1730.

\bibitem[Grossmann \& Lohse(2000)]{Grossmann} \textsc{Grossmann S.} and \textsc{Lohse, D.} 2000 Scaling in thermal convection: A unifying theory. \emph{J. Fluid Mech.} \textbf{407}, 27.

\bibitem[Hassanzadeh \etal (2014)]{Hassanzadeh} \textsc{Hassanzadeh P.}, \textsc{Chini G.P.} and \textsc{Doering C.R.} 2014 Wall to wall optimal transport. \emph{J. Fluid Mech.} \textbf{751}, 627-662.

\bibitem[Howard (1963)]{Howard} \textsc{Howard L.N.} 1963 Heat transport by turbulent convection. \emph{J. Fluid Mech.} \textbf{17}, 405.

\bibitem[Howard (1972)]{Howard} \textsc{Howard L.N.} 1972 Bounds on flow quantities. \emph{Ann. Rev. Fluid Mech.} \textbf{4}, 473-494.

\bibitem[Ierley \& Worthing (2001)]{Ierley} \textsc{Ierley G.R.} and \textsc{Worthing R.A.} 2001  Bound to improve: a variational approach to convective heat transport. \emph{J. Fluid Mech.} \textbf{441}, 223.

\bibitem[Kerswell (1998]{Kerswell} \textsc{Kerswell R.R.} 1998 Unification of variational principles for turbulent shear flows: the background method of Doering-Constantin and the mean-fluctuation formulation of Howard-Busse. \emph{Physica D} \textbf{121}, 175-192.

\bibitem[Kerswell (2001)]{Kerswell} \textsc{Kerswell R.R.} 2001 New results in the variational approach to turbulent Boussinesq convection. \emph{Phys. Fluids} \textbf{13}, 192.


\bibitem[Malkus (1954)]{Malkus} \textsc{Malkus W.V.R.} 1954 The heat transport and spectrum of thermal turbulence. \emph{Proc. R. Soc. London A} \textbf{225}, 196.

\bibitem[Motoki \etal (2018)]{Motoki} \textsc{Motoki S.}, \textsc{Kawahara G.} and \textsc{Shimizu M.} 2018 Maximal heat transfer between two parallel plates. \emph{J. Fluid Mech.} {\bf 851}, R4.


\bibitem[Plasting \& Kerswell (2003)]{Plasting1} \textsc{Plasting S.C.} and \textsc{Kerswell R.R.} 2003 Improved upper bound on the energy dissipation rate in plane Couette flow: the full solution to Busse's problem and the Constantin-Doering-Hopf problem with one-dimensional background field. \emph{J. Fluid Mech.} \textbf{16}, 363-379 ({\bf referred to as PK03 in the text}).

\bibitem[Plasting \& Kerswell (2005]{Plasting2} \textsc{Plasting S.C.} and \textsc{Kerswell R.R.} 2005 A friction factor bound for transitional pipe flow \emph{Phys. Fluids} \textbf{17}, 011706.

\bibitem[Priestley]{Priestley}\textsc{Priestley C.H.B} 1954 Convection from a large horizontal surface \emph{Aust. J. Phys.} \textbf{7}, 176-201 .


\bibitem[Sondak \etal (2015)]{Waleffe2} \textsc{Sondak D.}, \textsc{Smith L.} and \textsc{Waleffe F.} 2015 Optimal heat transport solutions for Rayleigh-B\'enard convection. \emph{J. Fluid Mech.} \textbf{784}, 565.

\bibitem[Souza (2016)]{Souza1} \textsc{Souza A.} 2016 An optimal control approach to bounding transport properties of thermal convection. \emph{Ph.D. thesis}, \emph{University of Michigan}.

\bibitem[Souza \etal (2019)]{Souza2} \textsc{Souza A.}, \textsc{Tobasco I.} and \textsc{Doering C.} 2019 Wall-to-wall optimal transport: theory and 2D computations. \emph{preprint}


\bibitem[Tobasco \& Doering (2017)]{Tobasco} \textsc{Tobasco I.} and \textsc{Doering C.R.} 2017 Optimal wall-to-wall transport by incompressible flows. \emph{Phys. Rev. Lett.} \textbf{118}, 264502.

\bibitem[Tobasco \etal (2018)]{TobGolDoer} \textsc{Tobasco I.} \textsc{ Goluskin, D.}  and \textsc{Doering C.R.} 2018 Optimal bounds and extremal trajectories for time averages in nonlinear dynamical systems. \emph{Phys. Lett. A} \textbf{382}, 382-386.

\bibitem[Waleffe \etal (2015)]{Waleffe1} \textsc{Waleffe F.}, \textsc{Boonkasame A.} and \textsc{Smith L.} 2015 Heat transport by coherent Rayleigh-B\'enard convection. \emph{Phys. Fluids} \textbf{27}, 051702.

\bibitem[Wen \etal \,(2013)]{Wen} \textsc{Wen B.}, \textsc{Chini G.P.}, \textsc{Dianati N.}and \textsc{Doering C.} 2013
Computational approaches to aspect-ratio-dependent upper bounds and
heat flux in porous medium convection \emph{Phys. Lett. A} \textbf{A 377}, 2931.

\bibitem[Wen \etal \,(2015)]{Wen} \textsc{Wen B.}, \textsc{Chini G.P.}, \textsc{Kerswell R.R.} and \textsc{Doering C.} 2015 Time-stepping approach for solving upper-bound problems: application to two-dimensional Rayleigh-B\'enard convection. \emph{Phys. Rev. E} \textbf{92}, 043012.

\bibitem[Whitehead \& Doering (2011)]{Whitehead1} \textsc{Whitehead J.} and \textsc{Doering C.R.} 2011 Ultimate state of two-dimensional Rayleigh-B\'enard convection between free-slip fixed-temperature boundaries. \emph{Phys. Rev. Lett.} \textbf{106}, 244501.

\bibitem[Whitehead \& Doering (2012)]{Whitehead2} \textsc{Whitehead J.} and \textsc{Doering C.R.} 2012 Rigid bounds on heat transport by a fluid between slippery boundaries. \emph{J. Fluid Mech.} \textbf{707}, 241.

\bibitem[Zhu \etal (2018)]{Zhu} \textsc{Zhu X.}, \textsc{Mathai V.}, \textsc{Stevens R.J.A.M.}, \textsc{ Verzicco R.} and \textsc{Lohse D.} 2018 Transition to the ultimate regime in two-dimensional Rayleigh-Benard convection. \emph{Phys. Rev. Lett.} \textbf{120}, 144502.



\end{thebibliography}

\end{document}